\newcommand{\arxivVersion}{\True}
\newcommand{\fillinbtw}{\False}

\documentclass[orivec,runningheads,letterpaper]{llncs} 

\usepackage{graphicx}
\usepackage[tight,footnotesize]{subfigure}
\usepackage{tikz}
\usetikzlibrary{arrows,chains,matrix,positioning,scopes}
\usepackage{amsmath}
\usepackage{amssymb}
\usepackage{enumerate}
\usepackage{bbm}
\usepackage{dsfont}
\usepackage{pdfpages}
\usepackage{pgfplots}
\usetikzlibrary{plotmarks}
\usepackage{multirow}
\usepackage[nodayofweek]{datetime}
\usepackage{stmaryrd}
\usepackage{wasysym}
\usepackage{makecell}

\usepackage[noend]{algorithmic}
\usepackage{algorithm}
\usepackage{url}

\newlength {\localLength}    

\newcommand{\newlengthsettowidth}[2]{\newlength {#1} \settowidth{#1}{#2}}
\newcommand{\newcounterset}      [2]{\newcounter{#1} \setcounter{#1}{#2}}

\newcommand{\ensurecommand}[2]{
  \providecommand{#1}{}
  \renewcommand{#1}{#2}}

\newcommand{\NN}{\mathds N}                             
\newcommand{\ZZ}{\mathds Z}                             

\newcommand{\range}[3][X]{\Ifthen{\Equal{#1}{X}}{\{}#2,\ldots,#3\Ifthen{\Equal{#1}{X}}{\}}} 

\newcommand{\atl}{\geq}                                 
\newcommand{\atm}{\leq}                                 

\newcommand{\union}       {\mathbin        {\cup}}      
\renewcommand {\subset}     {\subseteq}                 
\ensurecommand{\subsetneq}  {\subsetneqq}               

\newcommand{\func}[3]{#1 \colon #2 \rightarrow #3}      
\newcommand{\restr}[2]{#1\big|_{#2}}                    

\newcommand{\angles}  [1]{     \langle #1       \rangle} 

\newcommand{\true}    {\mathit{true}}

\newcommand{\Ifthenelse}[3]{\ifthenelse{#1}{#2}{#3}}   
\newcommand{\Ifthen}    [2]{\Ifthenelse{#1}{#2}{}}
\newcommand{\Unless}    [2]{\Ifthen{\not {#1}}{#2}}
\newcommand{\Equal}     [2]{\equal{#1}{#2}}            
\newcommand{\Empty}     [1]{\Equal{#1}{}}
\newcommand{\True}         {\Equal{1}{1}}
\newcommand{\False}        {\Equal{1}{2}}

\newdateformat{abbrevdate}{\shortmonthname\ \twodigit{\THEDAY}, \THEYEAR}

\providecommand{\Draftmode}{\False}

\newcommand{\Itedraft}    [2]{\Ifthenelse{\Draftmode}{#1}{#2}}
\newcommand{\Ifdraft}     [1]{\Itedraft{#1}{}}
\newcommand{\Unlessdraft} [1]{\Itedraft{}{#1}}
\newcommand{\drafttext}   [2][\draftcolor]{\Ifdraft{{\color{#1}#2}}}
\newcommand{\draftpointer}{\makebox[0mm][c]{$^*$}}
\newcommand{\draftmargin} [2][\draftcolor]{\drafttext[#1]{\draftpointer\marginpar[\raggedleft\small{\color{#1}#2}]{\raggedright\small{\color{#1}#2}}}} 
\newcommand{\draftdate}   [1][\abbrevdate\today, \currenttime]{\drafttext{\makebox[0mm][l]{\normalfont\tiny \ \ (Draft of #1)}}} 

\newcommand{\draftauthor} [3]{
  \newcommand{#1}[1]{\drafttext  [#3]{##1}}   
  \newcommand{#2}[1]{\draftmargin[#3]{##1}}}  

\newcommand{\useifspacecommand}[1]{\draftauthor{\ifspace}{\ifspacemargin}{#1}}

\Ifdraft{\setlength{\overfullrule}{10mm}}

\newcommand{\C}{\texttt C}

\newcommand{\centerTwoOut}  [2]               {#1 \hfill #2}                          

\newlength{\posLength}
\newcommand{\pos}[3][c]{\settowidth{\posLength}{#3}\makebox[\posLength][#1]{#2}}
\newlengthsettowidth{\tabLength}{\ \ \ }

\newcommand{\wbox}[2][\tabLength]{\hspace*{#1}\mbox{#2}\hspace*{#1}}

\newcommand{\Paragraph}[1][\baselineskip]{\vspace{#1}}
\newcommand{\0}[1]{}                                             
\newcommand{\End}{\end{document}}                                
\newcommand{\format}{}                                           
\newcommand{\emphasize}[1]{\textbf{#1}}                          
\newcommand{\emphdef}[1]{\emphasize{#1}}                         

\newcommand{\defFullOrAbbrev}[2]{#1} 

\newtheorem{DEF}     {\defFullOrAbbrev{Definition} {Def.}}
\newtheorem{THE}[DEF]{\defFullOrAbbrev{Theorem}    {Thm.}}
\newtheorem{LEM}[DEF]{\defFullOrAbbrev{Lemma}      {Lem.}}

\newtheorem{COR}[DEF]{\defFullOrAbbrev{Corollary}  {Cor.}}

\newcommand{\Proof}{\textbf{Proof}}                     
\newcommand{\explain}[2][\tab]{#1 \angles{\ \mbox{#2} \ }} 
\newcommand{\eop}[1][3mm]{\eopBox \vspace{#1}}          

\newcommand{\eopBox}{~\hfill$\Box$}                     

\algsetup{indent=1.5em}

\newcommand{\plkeyword}[1]{\textbf{#1}}

\newcommand{\FOREACH}{\FORALL}

\newcommand{\code}[1]{\texttt{#1}}

\newcommand{\assign}{\code{:=}}

 \algsetup{indent=.8em} 
\newcommand{\refCapitalOrSmall}[3]{#1#3} 

\newcommand{\refFullOrAbbrev}[2]{#1} 

\newcommand{\appendixref}   [2][!]{\genericref[#1] A a {\refFullOrAbbrev{ppendix}   {pp.}} {appendix}   {#2}}
\newcommand{\algorithmref}  [2][!]{\genericref[#1] A a {\refFullOrAbbrev{lgorithm}  {lg.}} {algorithm}  {#2}}

\newcommand{\definitionref} [2][!]{\genericref[#1] D d {\refFullOrAbbrev{efinition} {ef.}} {definition} {#2}}
\newcommand{\figureref}     [2][!]{\genericref[#1] F f {\refFullOrAbbrev{igure}     {ig.}} {figure}     {#2}}

\newcommand{\lemmaref}      [2][!]{\genericref[#1] L l {\refFullOrAbbrev{emma}      {em.}} {lemma}      {#2}}
\newcommand{\lineref}       [2][!]{\genericref[#1] L l {\refFullOrAbbrev{ine}       {ine}} {line}       {#2}}

\newcommand{\sectionref}    [2][!]{\genericref[#1] S s {\refFullOrAbbrev{ection}    {ect.}}{section}    {#2}}

\newcommand{\theoremref}    [2][!]{\genericref[#1] T t {\refFullOrAbbrev{heorem}    {hm.}} {theorem}    {#2}}

\newcommand{\equationref}[2][!]{\Ifthen{\Equal{#1}!}{\refCapitalOrSmall E e {\refFullOrAbbrev{quation}{q.}}~}(\ref{equation: #2})}

\newcommand{\genericref} [6][!]{\Ifthen{\Equal{#1}!}{\refCapitalOrSmall{#2}{#3}{#4}~}\ref{#5: #6}} 

\newcommand{\CASE}[1]{\STATE \textbf{case} #1\textbf{:} \begin{ALC@g}}
\newcommand{\ENDCASE}{\end{ALC@g}}

\newcommand{\DEFAULT}{\STATE \textbf{default:} \begin{ALC@g}}
\newcommand{\ENDDEFAULT}{\end{ALC@g}}
\newcommand{\DEFAULTLINE}[1]{\STATE \textbf{default:} }

\renewcommand{\refFullOrAbbrev}[2]{#2} 
\renewcommand{\defFullOrAbbrev}[2]{#2} 

\renewcommand{\bf}{\bfseries}

\renewcommand{\it}{\itshape}

\draftauthor{\thomas}{\thomasmargin}{green}
\draftauthor{\peizun}{\peizunmargin}{blue}

\newcommand{\ifspacecolor}{cyan}
\useifspacecommand{\ifspacecolor}
\Unlessdraft{\renewcommand{\ifspace}[1]{#1}}

\newcommand{\nonsimple}{spaghetti}
\newcommand{\Nonsimple}{Spaghetti}

\newcommand{\symbolcomment}[2]{\stackrel{\mbox{\tiny{(#1)}}}{#2}}

\newcommand{\expand}  [1]{{#1}^+}
\newcommand{\quotient}[1]{\overline{#1}}

\newcommand{\familyoperator}{_} 

\newcommand{\ttdsymbol}{\mathcal P} 

\newcommand{\familyobject}[2]{\Ifthenelse{\Empty{#1}}{#2}{{#2} \familyoperator {#1}}} 

\newcommand{\ttd}      [1][]{\familyobject{#1}{\ttdsymbol}} 
\newcommand{\ttdshareds} S
\newcommand{\ttdlocals}[1][]{\familyobject{#1}{L}}
\newcommand{\ttdstates}[1][]{\familyobject{#1}{V}}
\newcommand{\ttdtranss}[1][]{\familyobject{#1}{R}}
\newcommand{\localpart} W
\newcommand{\numlocals}{|L|}
\newcommand   {\edgesym}    {\rightarrow}
\newcommand{\expedgesym}{\dashrightarrow}

\newcommand{\threadstate}[2]{(#1,#2)}

\newcommand{\edge}   [5][]{\threadstate{#2}{#3} \Ifthenelse{\Empty{#1}}{   \edgesym}{\stackrel{#1}{   \edgesym}} \threadstate{#4}{#5}} 
\newcommand{\expedge}[5][]{\threadstate{#2}{#3} \Ifthenelse{\Empty{#1}}{\expedgesym}{\stackrel{#1}{\expedgesym}} \threadstate{#4}{#5}} 

\newcommand{\ttdexpand}      {\expand   \ttd}
\newcommand{\ttdtranssexpand}{\expand   \ttdtranss}
\newcommand{\ttdquotient}    {\quotient \ttd}

\newcommand{\subscriptobject}[2]{\Ifthenelse{\Empty{#1}}{#2}{{#2}_{#1}}} 

\newcommand{\ttdshared}[1][]{\subscriptobject{#1}{s}}
\newcommand{\ttdlocal} [1][]{\subscriptobject{#1}{l}}
\newcommand{\ttdstate} [1][]{\subscriptobject{#1}{t}}

\newcommand{\ttdinitindex}{I}
\newcommand{\ttdinitshared}  {\ttdshared[\ttdinitindex]}
\newcommand{\ttdinitlocal}   {\ttdlocal [\ttdinitindex]}
\newcommand{\ttdinitstate}{\ttdstate [\ttdinitindex]}
\newcommand{\ttdinitstatepair}{\threadstate \ttdinitshared \ttdinitlocal}

\newcommand{\ttdfinalindex}{F}
\newcommand{\ttdfinalshared}  {\ttdshared[\ttdfinalindex]}
\newcommand{\ttdfinallocal}   {\ttdlocal [\ttdfinalindex]}
\newcommand{\ttdfinalstate}{\ttdstate [\ttdfinalindex]}
\newcommand{\ttdfinalstatepair}{\threadstate \ttdfinalshared \ttdfinallocal}

\newcommand{\ttdinf}{\ttd[\infty]}
\newcommand{\states}{\ttdstates[\infty]}
\newcommand{\transs}{\ttdtranss[\infty]}
\newcommand{\transsymbol} \rightarrowtail

\newcommand{\state}[2]{(#1|#2)} 

\newcommand{\covers}{\succeq}

\newcommand{\BWS}   {\textsc{Bws}}
\newcommand{\CovPre}{\textsc{CovPre}}

\newcommand{\ourtool}    {\textsc{Cutr}} 
\newcommand{\ourtoolurl} {\url{http://www.ccs.neu.edu/home/lpzun/cutr}}

\newcommand{\upper}{\operatorname{\uparrow}}

\newcommand{\maxplus}   [1][]{\mathbin{\varoplus}  \Unless{\Empty{#1}}{_{#1}}}
\newcommand{\maxminus}  [1][]{\mathbin{\varominus} \Unless{\Empty{#1}}{_{#1}}}

\newcommand{\iterator}{\kappa}
\newcommand{\iterate}[2][(\iterator)]{{#2}^{#1}} 

\newcommand{\Tau}{\mathcal T}

\begin{document}

\mainmatter  





\newcommand{\theTitle}{%
  Concolic Unbounded-Thread Reachability \\
  via Loop Summaries\\ 
  \Ifthen{\arxivVersion}{ (Extended Technical Report)}}

\titlerunning{Concolic Unbounded-Thread Reachability via Loop Summaries}

\title{\theTitle\thanks{This work is supported by US National Science Foundation grant no.\ 1253331.}}

%
%
\author{Peizun Liu
  \and Thomas Wahl\draftdate%
}

\authorrunning{Peizun Liu and Thomas Wahl}

\institute{Northeastern University, Boston, United States}

%
%
\toctitle{Peizun Liu's LNCS Template}
\tocauthor{Peizun Liu based on LNCS}
\maketitle


\begin{abstract}
  We present a method for accelerating explicit-state backward search
  algorithms for systems of arbitrarily many finite-state threads. Our
  method statically analyzes the program executed by the threads for the
  existence of simple loops. We show how such
  loops can be collapsed \emph{without approximation} into Presburger
  arithmetic constraints
  that symbolically summarize the effect of executing the backward search
  algorithm along the loop in the multi-threaded program.
  As a result, the subsequent explicit-state search does not need to
  explore the summarized part of the state space. The combination of
  concrete and symbolic exploration gives our algorithm a \emph{concolic}
  flavor. We demonstrate the power of this method for proving and refuting
  safety properties of unbounded-thread programs.
\end{abstract}



\section{Introduction}
\label{section: Introduction}

Unbounded-thread program verification continues to attract the attention it
deserves: it targets programs designed to run on multi-user platforms and
web servers, where concurrent software threads respond to service requests
of a number of clients that can usually neither be predicted nor
meaningfully bounded from above a priori. Such programs are therefore
designed for an unspecified and unbounded number of parallel threads as a
system parameter.

We target in this paper unbounded-thread shared-memory programs where each
thread executes a non-recursive Boolean (finite-data) procedure. This model
is popular, as it connects to multi-threaded \C\ programs via predicate
abstraction~\cite{GS97,BMMR01}, a technique that has enjoyed progress for concurrent programs
in recent years \cite{DKKW2011}. The model is also popular since basic
program state reachability questions are decidable. They are also, however,
of high complexity: the equivalent \emph{coverability problem} for Petri
nets was shown to be EXPSPACE hard~\cite{CLM76}. The motivation for our
work is therefore
to improve the efficiency of existing algorithms.

A sound and complete method for \emph{coverability} analysis for \emph{well
  quasi-ordered systems} (WQOS) is the backward search algorithm by Abdulla
\cite{A10}. Coverability for WQOS subsumes program state
reachability analysis for a wide class of multi-threaded Boolean programs.
Starting from the target state whose reachability is under investigation,
the algorithm proceeds backward by computing \emph{cover preimages},
until either an initial state is reached, or a fixpoint. This search
principle is used in several variants, such as the widening-based approach
in~\cite{KKW14a}.

In this paper we propose an idea to accelerate backward search algorithms
like Abdulla's. The goal is to symbolically \emph{summarize} parts of the
finite-state transition system $\ttd$ (our formal model for Boolean
programs) executed by each thread, in a way that reachability in the
summarized parts can be reduced to satisfiability of the summary formulas.
Prime candidates for such symbolic summaries are \emph{loops} in $\ttd$.
The exploration algorithm may have to traverse them multiple times before a
loop fixpoint is reached. We instead wish to summarize the loop statically,
obtaining a formula parameterized by the number $\iterator$ of loop
iterations, for the global state reached after $\iterator$ traversals of
the loop.

In order to enable loop summarization, our approach first builds an
abstraction $\ttdquotient$ of the transition graph $\ttd$ (i) that is
\emph{acyclic}, and (ii) whose \emph{single-threaded} execution
overapproximates the execution of $\ttd$ by \emph{any} number of threads.
Thus, if there is no single-threaded path to the final state in
$\ttdquotient$, the algorithm returns ``unreachable'' immediately.
Otherwise, since $\ttdquotient$ is acyclic, there are only finitely many
paths that require investigation.

For each such path, we now determine whether it is ``summarizable''. This
is the case if the path either features no loops, or only \emph{simple}
loops: single cyclic paths without nesting. We show in this paper how a
\emph{precise} summary of the execution of standard backward search across
such a path can be obtained as a formula in Presburger arithmetic, the
decidable theory over linear integer operations. Conjoined with appropriate
constraints encoding the symbolic initial and final states, reachability is
then equivalent to the satisfiability of this summary.

Our algorithm can be viewed as separating the branching required in the
explicit-state traversal in Abdulla's algorithm~\cite{A10}, and the
arithmetic required to keep track (via counting) of the threads in various
local states. Structure $\ttdquotient$ is loop-free and can thus be
explored path by path. Paths with only simple loops are symbolically
summarized into a Presburger formula. The question whether the target state
is reachable along this path can then often be answered quickly, in
part since the formulas tend to be easy to decide.
Other parts are explored using standard explicit-state traversal,
restricted to the narrow slice of the state space laid out by this path,
which gives our algorithm a \emph{concolic} flavor.

We conclude this paper with experiments that investigate the performance
gain of our acceleration method applied to backward search. The results
demonstrate that transition systems obtained from Boolean programs, which
feature ``execution discipline'' enforced by the control flow, are better
suited to pathwise acceleration than Petri nets, which often encode
rule-based (rather than program-based) transition systems and thus feature
fewer summarizable paths.

Proofs to claims made in this paper can be found in the
Appendix\Unless{\arxivVersion}{ of \cite{LW16b}}.

\section{Thread-Transition Diagrams and Backward Search}
\label{section: Thread-Transition Diagrams and Backward Search}

We assume multi-threaded programs are given in the form of an abstract
state machine called \emph{thread transition diagram} \cite{KKW14a}. Such a
diagram
reflects the replicated nature of programs we consider: programs consisting
of threads executing a given procedure defined over shared (``global'') and
(procedure-)local variables. A~thread transition diagram (TTD) is a tuple
$\ttd=(\ttdshareds,\ttdlocals,\ttdtranss)$, where
\begin{itemize}
  \item $\ttdshareds$ is a finite set of \emph{shared} states;
  \item $\ttdlocals$ is a finite set of \emph{local} states;
  \item $\ttdtranss \subset (\ttdshareds \times \ttdlocals) \times (\ttdshareds \times
    \ttdlocals)$ is a (finite) set of \emph{edges}.
\end{itemize}
An element of $\ttdstates = \ttdshareds \times \ttdlocals$ is called
\emph{thread state}. We write $\edge{s_1}{l_1}{s_2}{l_2}$ for
$((s_1,l_1),(s_2,l_2)) \in \ttdtranss$. We assume the TTD has a
\emph{unique} initial thread state, denoted $\ttdinitstate =
\ttdinitstatepair$; the case of multiple initial thread states is discussed
in \Ifthenelse{\arxivVersion}{\appendixref{Uniqueness of the Initial
    State}}{App. A}\Unless{\arxivVersion}{ of
  \cite{LW16b}}.
An example of a TTD is shown in \figureref{ttd}.
\begin{figure}[htbp]
  \centerline{
    \subfigure[]{\includegraphics[width=1.6in]{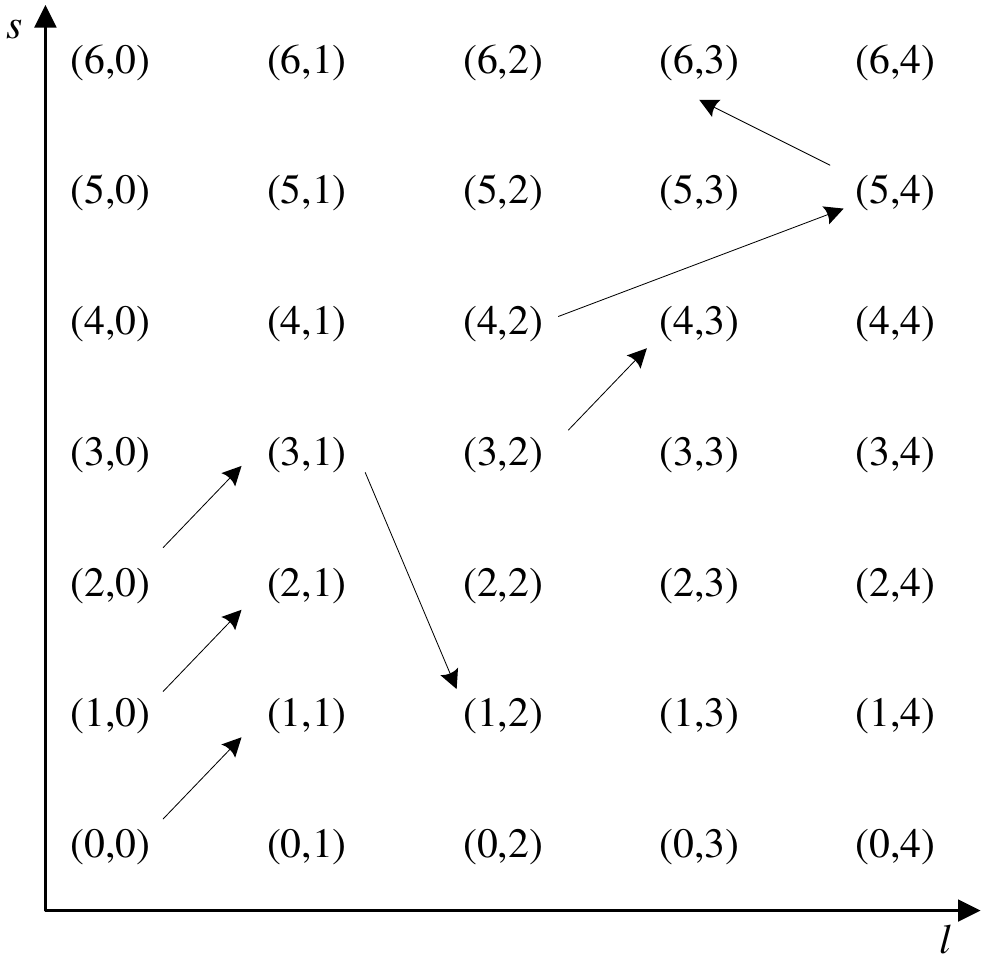}
      \label{figure: ttd}}
    \hfil
    \hspace{0.2em}
    \subfigure[]{\includegraphics[width=1.6in]{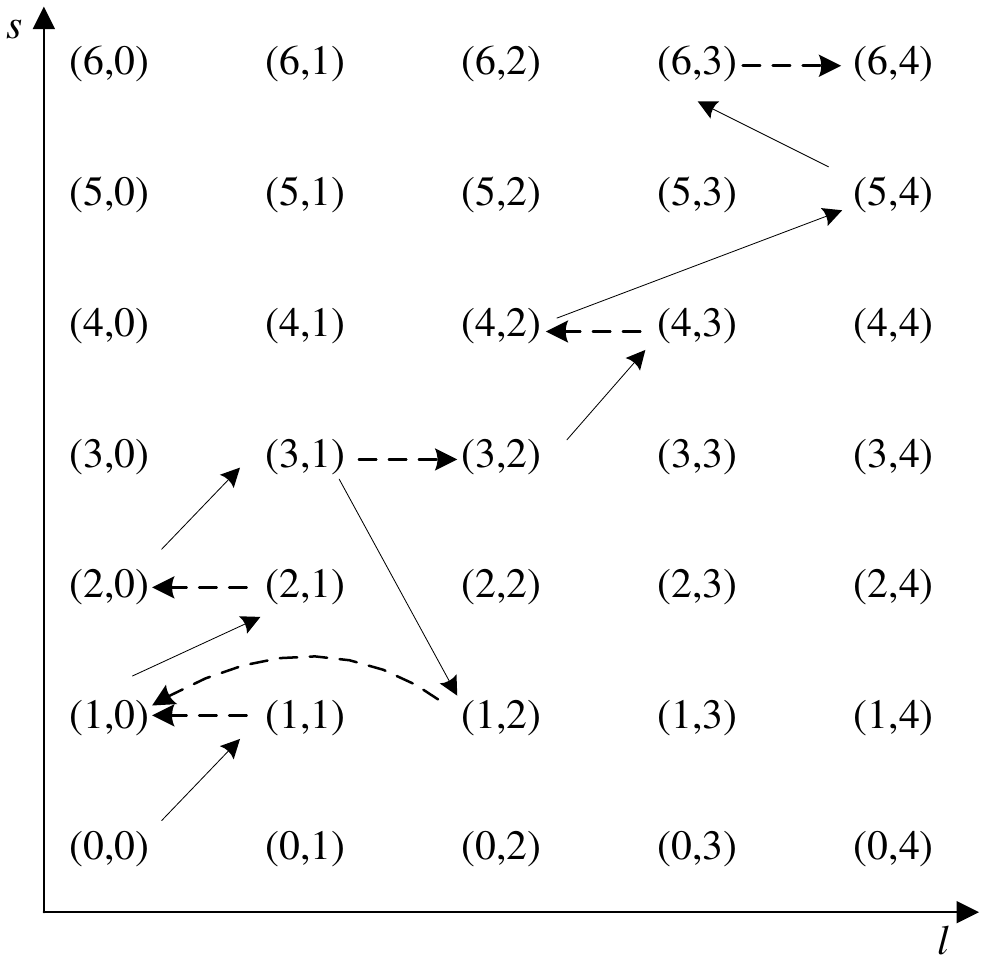}
      \label{figure: expanded ttd}}
    \hfil
    \hspace{0.2em}
    \subfigure[]{\includegraphics[width=1.35in]{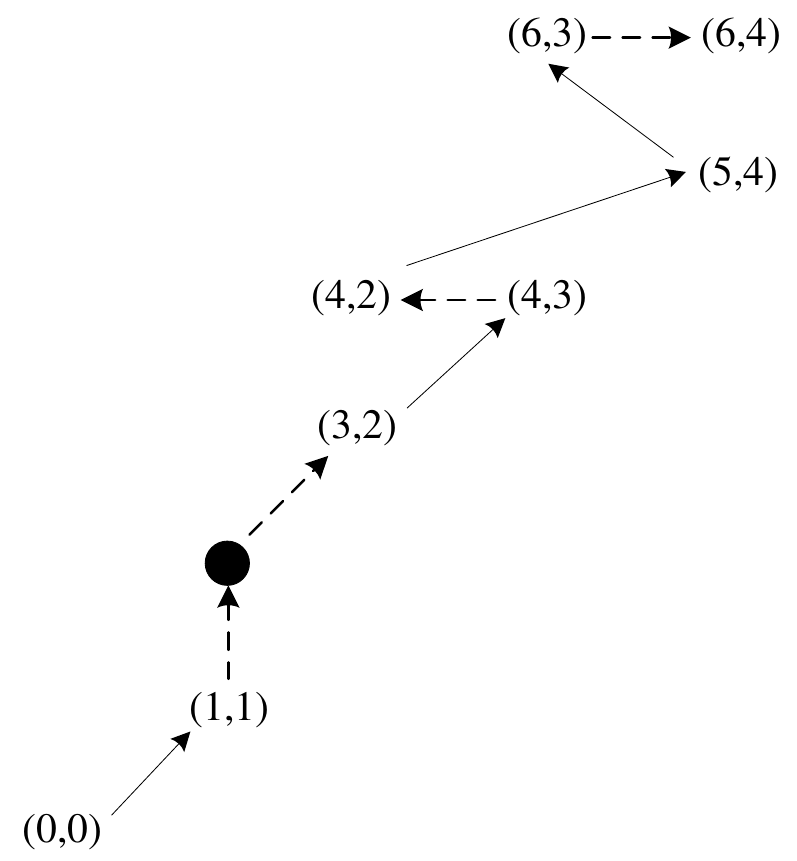}
      \label{figure: quotient ttd}}}
  \caption{(a) A thread transition diagram $\ttd$ (initial state
    $\ttdinitstate = \threadstate 0 0$); (b) the Expanded TTD $\ttdexpand$
    with a path~$\expand \sigma$; (c) the SCC quotient graph $\ttdquotient$
    of $\ttdexpand$, with quotient path $\quotient \sigma$. The black disc
    represents the loop in~$\expand \sigma$ (the other SCCs are trivial)}
  \label{figure: ttd abstractions}
\end{figure}

A TTD gives rise to a family, parameterized by $n$, of transition systems
$\ttd[n] = (\ttdstates[n],\ttdtranss[n])$ over the state space
$\ttdstates[n] = \ttdshareds \times \ttdlocals^n$, whose states we write in
the form $\state s {\range[]{l_1}{l_n}}$. This notation represents a global
system state with shared component $s$, and $n$ threads in local states
$l_i$, for $i \in \range 1 n$.
The transitions of $\ttd[n]$, forming the set $\ttdtranss[n]$, are written
in the form $\state s {\range[]{l_1}{l_n}} \transsymbol \state {s'}
{l'_1,\ldots,l'_n}$. This transition is defined exactly if there exists $i
\in \range 1 n$ such that $\edge s {l_i} {s'} {l'_i}$ and for all $j \not=
i$, $l_j = l'_j$. That is, our execution model is asynchronous: each
transition affects the local state of at most one
thread.\ifspace{\footnote{\emph{Dynamic thread creation} is discussed at
    end of \sectionref{Pathwise Unbounded-Thread Reachability}}}

The initial state set of $\ttd[n]$ is $\{\ttdinitshared\} \times
\{\ttdinitlocal\}^n$. A \emph{path} of $\ttd[n]$ is a finite sequence of
states in $\ttdstates[n]$ whose first element is initial, and whose
adjacent elements are related by $\ttdtranss[n]$. A thread state $(s,l) \in
\ttdshareds \times \ttdlocals$ is \emph{reachable in $\ttd[n]$} if there
exists a path in $\ttd[n]$ ending in a state with shared state $s$ and some
thread in local state~$l$.

A TTD also gives rise to an infinite-state transition system $\ttdinf =
(\states,\transs)$, whose set of states/transitions/initial states/paths is
the union of the sets of states/transitions/initial states/paths of
$\ttd[n]$, for all $n \in \NN$. We are tackling in this paper the
\emph{thread state reachability problem}: given a TTD $\ttd$ and a
\emph{final} thread state $\ttdfinalstatepair$, is $\ttdfinalstatepair$
reachable in $\ttdinf$~? It is easy to show that this question is
decidable, by reducing $\ttdinf$ to a \emph{well quasi-ordered system}
(WQOS)~\cite{A10}: let the \emph{covers} relation $\covers$ over
$\states$ be defined as follows:
\[
  \state s {l_1,\ldots,l_n} \covers \state {s'} {l'_1,\ldots,l'_{n'}}
\]
whenever $s = s'$ and for all $l \in L$, $|\{i: l_i = l\}| \atl |\{i: l'_i
= l\}|$. The latter inequality states that the number of threads in local
state $l$ ``on the left'' is at least the number of threads in local state
$l$ ``on the right''.
Relation $\covers$ is a well quasi-order on $\states$, and
\mbox{$(\ttdinf,\covers)$} satisfies the definition of a WQOS, in
particular the \emph{monotonicity} property required of $\covers$ and
$\transsymbol$. The proof of this property exploits the \emph{symmetry} of
the multi-threaded system: the threads execute the same program $\ttd$: a
state $\state s {\range[]{l_1}{l_n}}$ can be compressed without loss of
information into the counter notation $\state s
{\range[]{n_1}{n_{\numlocals}}}$, where $n_l = |\{i: l_i = l\}|$.

The thread state reachability question can now be cast as a
\emph{coverability problem}, which is decidable but of high complexity,
e.g.\ EXPSPACE-hard for standard Petri nets \cite{CLM76}, which are
equivalent in expressiveness to infinite-state transition systems obtained
from TTD \cite{KKW14a}.


\noindent
\setlength{\localLength}{\parindent} 
\begin{minipage}{70mm}
  \hspace*{\localLength}A sound and complete algorithm to decide
  coverability for WQOS is the \emph{backward search} algorithm by Abdulla
  et al.\ \mbox{\cite{ACJT96,A10}}, a simple version of which is
  shown on the right. Input is a WQOS $M$, a set of initial states $I$, and
  a non-initial final state~$q$. The algorithm maintains a work set $W$ of
  unprocessed states, and a set $U$ of minimal encountered states. It
  iteratively computes minimal \emph{cover predecessors}
  \begin{equation}
    \format\hspace*{-9pt}\CovPre(w) = \min\{p : \exists w' \covers w : p \transsymbol w'\}
    \label{equation: cover predecessors}
  \end{equation}
  starting from $q$, and terminates either by back\format\-ward-reaching an
  initial state (thus proving coverability of $q$), or when no unprocessed
  vertex remains (thus proving uncoverability).
\end{minipage}
\hfill
\begin{minipage}{46mm}
  \vspace*{-7mm}
  \begin{algorithm}[H]
    \caption{$\BWS(M,I,q)$}
    \begin{algorithmic}[1]
      \REQUIRE{initial states $I$, \\ \quad \quad final state $q \not\in I$}
      \STMT $W := \{q\}$\,; \,$U := \{q\}$
      \WHILE{$\exists w \in W$}
        \STMT $W := W \setminus \{w\}$
        \FOR{$p \in \CovPre(w) \setminus \upper U$} \label{line: bwpreimage}
          \IF{$p \in I$}
            \STMT ``$q$ coverable''
          \ENDIF
          \STMT $W := \min(W \union \{p\})$ \label{line: discard non minimals 1}
          \STMT $U := \min(U \union \{p\})$ \label{line: discard non minimals 2}
        \ENDFOR
      \ENDWHILE
      \STMT ``$q$ not coverable'' \label{line: bc: uncoverable}
    \end{algorithmic}
    \label{algorithm: Abdulla}
  \end{algorithm}
  \vspace{-6mm}
  \footnotesize
  \algorithmref{Abdulla}: infinite-state backward search. Symbol $\upper U$
  stands for the \emph{upward closure} of $U$: \\
  $\upper U = \{u': \exists u \in U: u' \covers u\}.$
\end{minipage}

\ifspace{\paragraph{Strongly connected components.} In this paper we also
  frequently make use of the following standard notions. Given a directed
  graph $G$, a \emph{strongly connected component} (SCC) is a maximal set
  $C$ of vertices such that for any two vertices $c_1$ and $c_2$ in $C$,
  there is a path in $C$ from $c_1$ to $c_2$. If the subgraph of $G$
  induced by $C$ has no edge, $C$ is called \emph{trivial}.

  The \emph{SCC quotient graph} $\quotient G$ of $G$ has exactly one vertex
  for each SCC of $G$, and no other vertices; we identify each vertex of
  $\quotient G$ with the SCC it represents. An edge $(C_1,C_2)$ exists in
  $\quotient G$ whenever $C_1 \not= C_2$ and there is a $G$-edge from some
  vertex in $C_1$ to some vertex in $C_2$. For a vertex $v$, we denote by
  $\quotient v$ the unique SCC that $v$ belongs to (hence, by
  identification, $\quotient v$ is also a vertex in $\quotient G$). Since
  each cycle of $G$ is contained entirely in one SCC, and nodes in
  $\quotient G$ have no self-loops, $\quotient G$ is \emph{acyclic}.}

\section{Pathwise Unbounded-Thread Reachability: Overview}
\label{section: Pathwise Unbounded-Thread Reachability: Overview}

Our approach for accelerating backward reachability analysis is two-phased.
The first phase constructs from $\ttd$ an abstract structure
$\ttdquotient$, with the property that any thread state reachable in
$\ttdinf$ (i.e., for any number of threads) is also reachable in
$\ttdquotient$ \emphasize{when executed by a single thread}. Structure
$\ttdquotient$ thus overapproximates the thread-state reachability problem
for $\ttd$ to a much simpler sequential reachability problem. Technically,
the abstraction first adds certain edges to~$\ttd$,
and then collapses strongly connected components
to obtain $\ttdquotient$, which is hence acyclic.
Note that this first phase performs no exploration and is in fact
independent of the underlying reachability algorithm being accelerated.

In the second phase, we analyze each path $\quotient \sigma$ in the acyclic
structure~$\ttdquotient$ from $\ttdinitstate$ to $\ttdfinalstate$
separately, if any. We now distinguish: if $\quotient \sigma$ visits only
\emph{simple} SCCs, by which we mean SCCs that represent simple loops, then
we call $\quotient \sigma$ simple, and we precisely summarize the effect of
traversing the path using Presburger formulas.\footnote{Simple SCC nodes
  (representing a simple loop) are not to be confused with \emph{trivial}
  SCC nodes (representing a single node). Simple nodes are by definition
  non-trivial} Instead of executing \algorithmref{Abdulla}, we solve these
Presburger constraints, in effect accelerating the algorithm, losslessly,
along loop-free path segments and simple loops. If $\quotient \sigma$
visits at least one \emph{\nonsimple\ SCC} --- an SCC that represents more
than a simple loop (e.g.\ a loop nest) --- then we call $\quotient \sigma$
\emph{\nonsimple} as well and explore it using \algorithmref{Abdulla},
restricted to the edges along $\quotient \sigma$.

At the end of this section we illustrate the overall process in more
detail. We first introduce the acyclic quotient structure $\ttdquotient$.

\subsubsection{A Single-Threaded Abstraction of $\ttdinf$.}
\label{section: A Single-Threaded Abstraction of ttdinf}

A key operation employed during backward search is what we call
\emph{expansion} of a global state: the addition of a thread in a suitable
local state during the computation of the cover
preimage~\equationref[]{cover predecessors}. We can simulate the effect of
such expansions \emph{without adding threads}, by allowing a thread to
change its local state in certain disciplined ways. To this end, we expand
the TTD data structure as follows.
\begin{DEF}
  \label{definition: expanded ttd}
  Given a TTD $\ttd=(\ttdshareds,\ttdlocals,\ttdtranss)$, an
  \emphdef{expansion edge} is an edge of the form $(\threadstate s l,
  \threadstate s {l'})$ (same shared state) such that $l \not= l'$ and the
  following holds:
  \begin{itemize}

  \item there exists an edge of the form \ $\ldots \edgesym \threadstate s
    l$ in $\ttdtranss$, \emphasize{and}

  \item there exists an edge of the form \ $\threadstate s {l'} \edgesym
    \ldots$ in $\ttdtranss$, or $(s,l') = \ttdfinalstatepair$.

  \end{itemize}
  The \emphdef{Expanded TTD (ETTD)} of $\ttd$ is the structure
  $\ttdexpand=(\ttdshareds,\ttdlocals,\ttdtranssexpand)$ with
  $\ttdtranssexpand = \ttdtranss \cup \{e: \ \mbox{$e$ is an expansion edge}\}$.
\end{DEF}
To distinguish the edge types in $\ttdexpand$, we speak of \emph{real
  edges} (in $\ttdtranss$) and expansion edges. Intuitively, expansion
edges close the gap between two real edges whose target and source,
respectively, differ only in the local state. This can be seen in
\figureref{expanded ttd}, which shows the ETTD generated from the TTD in
\figureref{ttd}. In the graphical representation, expansion edges run
horizontally and are shown as dashed arrows $\expedge s l s {l'}$.

To facilitate the identification and treatment of loops, we collapse the
ETTD $\ttdexpand$ into its (acyclic) SCC quotient graph, denoted
$\ttdquotient$. An example is shown in \figureref{quotient ttd}. For ease
of presentation, we assume that both the initial and final thread states
$\ttdinitstate$ and $\ttdfinalstate$ of $\ttd$ form single-node SCCs in
$\ttdquotient$, i.e.\ loops occur only in the interior of a path. This can
be enforced easily using artificial states.

Being acyclic, the quotient graph $\ttdquotient$ contains only finitely
many paths between any two nodes. It also has another key property that
makes it attractive for our approach. Let us interpret
$\ttdquotient$ as a sequential transition system. That is, when we speak
of \emph{reachability of a thread state} and \emph{paths} in
$\ttdquotient$, we assume
$\ttdquotient$ is executed by a single thread from $\ttdinitstate$. (In
contrast, the semantics of $\ttd$ is defined via the unbounded-thread
transition system $\ttdinf$.) Given these stipulations,
$\ttdquotient$~overapproximates $\ttd$, in the following sense:
\newcounterset{lemmaCounterValues}{\theDEF}
\begin{LEM}
  \label{lemma: quotient structure overapproximates}
  If thread state $\ttdfinalstate$ is reachable in $\ttdinf$, then
  $\ttdfinalstate$ is also reachable in $\ttdquotient$.
\end{LEM}

By \lemmaref{quotient structure overapproximates}, if $\ttdfinalstate$ is
not reachable from $\ttdinitstate$ in $\ttdquotient$ (a simple sequential
reachability problem), it is not reachable in $\ttdinf$. In that case our
algorithm immediately returns ``unreachable'' and terminates. If
$\ttdfinalstate$ is reachable in $\ttdquotient$, we cannot conclude
reachability in $\ttdinf$, as can be seen from \figureref{ttd
  abstractions}: thread state $\ttdfinalstate := \threadstate 6 4$ is
easily seen to be unreachable in $\ttdinf$, no matter how many threads
execute the diagram $\ttd$ in (a). But $\ttdfinalstate$ is obviously
sequentially reachable in $\ttdquotient$ (c). In the rest of this paper we
describe how to decide, for each path $\quotient \sigma$ in $\ttdquotient$
from $\ttdinitstate$ to $\ttdfinalstate$, whether it actually witnesses
reachability of $\ttdfinalstate$ in $\ttdinf$.

\Paragraph

To give an overview of this process, consider a quotient path $\quotient
\sigma$ with one simple SCC node. One such path is schematically depicted
in \figureref{path decomposition}, where we have zoomed in on the SCC node
$\ell_i$ in order to show the simple loop of $\ttdexpand$ collapsed inside
it. To analyze reachability of $\ttdfinalstate$ in $\ttdinf$, we first
consider the path segment from $\ttdfinalstate$ to the \emph{exit point} of
the loop (see \figureref{path decomposition}). The exit point is the unique
node of $\ttdexpand$ abstracted by SCC node $\ell_i$ that is first
encountered when the quotient path $\quotient \sigma$ is explored
\emph{backward}.
\begin{figure}[htbp]
  \centering
  {\begin{minipage}{100mm}
    \tikzstyle{selected} = [draw,line width=5pt,-,black!22,shorten >=0pt]
    \tikzstyle{null}=[fill=white!20,minimum size=0pt, inner sep=0pt]
    \tikzstyle{node}=[fill=white!20, inner sep=0pt]
    \tikzstyle{main node}=[circle,fill=white!20,draw,font=\sffamily\bfseries]
    \tikzstyle{tiny node}=[circle,fill=black!50,draw, minimum size=3pt, inner sep=0pt]
    \begin{tikzpicture}[->,auto,node distance=2.0cm,
        thick, scale=0.75, transform shape]
      \begin{scope}[xshift=0cm]
        \node[node] (0) [label=below:$\ttdinitstate$]{};
        \node[main node] (1) [above left of=0, label=above:$\ell_1$]{};
        \node[main node] (2) [below left of=1, label=below:$\ell_{i-1}$] {};
        \node[tiny node] (3) [above left of=2, label=above:] {};
        \node[null] (4) [above left of=3, label=above:] {};
        \node[null] (5) [below left of=3, label=below:] {};
        \node[null] (12) [above = 10mm of 5] {\Huge $\ell_i$};
        \node[tiny node] (6) [below left of=4, label=below:] {};
        \node[main node] (7) [below left of=6, label=below:$\ell_{i+1}$] {};
        \node[main node] (8) [above left of=7, label=above:$\ell_{m-1}$] {};
        \node[main node] (9) [below left of=8, label=below:$\ell_m$] {};
        \node[node] (10) [above left of=9, label=above:$\ttdfinalstate$]{};

        \path (0) edge             node{$h_0$} (1)
              (1) edge [dotted]    node{} (2)
              (7) edge [dotted]    node{$$} (8)
              (8) edge             node{$h_{m-1}$} (9)
              (9) edge             node{$h_m$} (10)
              (2) edge             node{} (3)
              (6) edge             node[label=above:$h_i$]{} (7)
              (5) edge [bend left=38] (6)
              (3) edge [bend left=38] node{$h_{i-1}$} (5)
              (4) edge [bend left=38] (3)
              (6) edge [bend left=38] (4);

        \draw[>=latex,thin](-7.3,2.5)node[left,scale=1.3, inner sep=1pt]{\emph{exit point}}
              to[out=0,in=160] (6);
        \draw[>=latex,thin](-4.0,2.5)node[right,scale=1.3, inner sep=1pt]{\emph{entry point}}
              to[out=180,in=20] (3);
      \end{scope}
    \end{tikzpicture}
  \end{minipage}}
  \caption{A path $\quotient \sigma$ in the acyclic structure
    $\ttdquotient$ with a non-trivial and magnified SCC node~$\ell_i$,
    representing some kind of loop structure in $\ttdexpand$}
  \label{figure: path decomposition}
\end{figure}

Our approach builds a symbolic summary for this path segment. We then do
the same for the simple loop collapsed inside $\ell_i$, and for the path
from the \emph{entry point} of the loop back to $\ttdinitstate$. These
summaries are combined conjunctively into a single Presburger expression
$\varphi$ over a parameter $\iterator$ that represents the number of
iterations through the loop represented by~$\ell_i$. We now conjoin
$\varphi$ with the constraint that, when backward-reaching $\ttdinitstate$
along~$\quotient \sigma$, no thread resides in any local state \emph{other
  than} $\ttdinitlocal$. This condition ensures that the global state
constructed via symbolic backward execution is of the form
$\{\ttdinitshared\} \times \{\ttdinitlocal\}^n$, i.e.\ it is initial. The
claim that $\ttdfinalstate$ is reachable in $\ttdinf$ is then equivalent to
the satisfiability of the overall formula; a satisfying assignment to
$\iterator$ specifies how many times the loop in $\ell_i$ needs to be
traversed.

\Paragraph

In \sectionref{Presburger Summaries for Loop-Free Path Segments}
and~\sectionref[]{Presburger Summaries for Simple Loops} we describe how
loop-free path segments and simple loops, respectively, are summarized, to
obtain a symbolic characterization.

\section{Presburger Summaries for Loop-Free Path Segments}
\label{section: Presburger Summaries for Loop-Free Path Segments}

Consider a path segment $\quotient \sigma$ in $\ttdquotient$ with only
trivial (singleton) SCC nodes in its interior; we call such segments
\emph{loop-free}. (The start and end state of $\quotient \sigma$ may still
be non-trivial SCC nodes; the loops contracted by these SCC nodes are not
considered in this section.) The real and expansion edges along $\quotient
\sigma$ suggest a \emph{firing sequence} of edges during an exploration of
$\ttdinf$ using \algorithmref{Abdulla}. Each real edge corresponds to a
thread state change for a single thread; each expansion edge corresponds to
the expansion of the current global state. More precisely, given a global
state of the form $\state{s'}{\range[]{l'_1}{l'_n}}$,
\algorithmref{Abdulla} computes cover preimages (\equationref{cover
  predecessors}), by first firing edges of $\ttdtranss$ backward whose
targets equal one of the thread states $\threadstate{s'}{l'_i}$. Second,
for each edge $e$ whose target $\threadstate{s'}{l'}$ (with shared state
$s'$) does not match any of the thread states $\threadstate{s'}{l'_i}$,
\algorithmref{Abdulla} expands the global state, by adding one thread in
local state $l'$, followed by firing $e$ backward, using the added
thread.\footnote{We exploit the fact that cover preimages in systems
  induced by TTDs increase the number of threads in a state by at most 1
  (see \cite[Lemma~1]{LW14} for a proof)}

The steps performed by \algorithmref{Abdulla} can be expressed in terms of
updates to local-state counters. Let edge $e$ be of the form $\edge s l
{s'}{l'}$. If the current global state $\state{s'}{\range[]{l'_1}{l'_n}}$
contains a thread in local state $l'$, firing $e$ backward amounts to
decrementing the counter $n_{l'}$ for the target $l'$, and incrementing the
counter $n_l$ for the source $l$. If the current global state does not
contain a thread in local state $l'$, we first expand the state by adding
such a thread, followed by firing $e$ backward. Together the step amounts
exactly to an increment of $n_l$.

We can execute these steps \emph{symbolically}, instead of concretely, by
traversing path segment $\quotient \sigma$ backward and encoding the
corresponding counter updates described in the previous paragraph as
logical constraints over the local-state counters. The constraints are
expressible in \emph{Presburger} (linear integer) arithmetic. To
demonstrate this, we introduce some light notation. For $x,y \in \ZZ$ and
$b \in \NN$, let $x \maxplus[b] y = \max\{x + y,b\}$. Intuitively, $x
\maxplus[b] y$ is ``$x + y$ but at least $b$''. When $b=0$, we omit the
subscript. We also use $x \maxminus[b] y$ as a shorthand for $x \maxplus[b]
(-y)$ ($ = \max\{x-y,b\}$). For example, $x \maxminus 1$ equals $x-1$ if $x
\atl 1$, and 0 otherwise.
Neither $\maxplus[b]$ nor $\maxminus[b]$ are associative: $(1 \maxplus 2)
\maxplus -3 = 0 \not= 1 = 1 \maxplus (2 \maxplus -3)$. We therefore
stipulate: these operators associate from left to right, and they have the
same binding power as $+$ and $-$ .

Operators $\maxplus/\maxminus$ in Presburger formulas are syntactic sugar:
we can rewrite a formula $\Gamma$ containing $x \maxplus[b] y$, using a
fresh variable $v$ per occurrence:
\begin{equation}
  \label{equation: maxplus rewriting}
  \Gamma \wbox{$\equiv$} (\Gamma|_{(x \maxplus[b] y) \rightarrow v}) \ \land \ ((x + y \geq b \land v = x + y) \lor (x + y < b \land v = b))
\end{equation}
where $\alpha|_{\beta \rightarrow \gamma}$ denotes substitution of $\gamma$
for $\beta$ in $\alpha$.

The \emph{summary} of loop-free path segment $\quotient \sigma$ is computed
separately for each local state $\ttdlocal$: \algorithmref{Symbolically
  executing a path for local state l} symbolically executes $\quotient
\sigma$ backward; for certain edges a ``contribution'' to counter
$n_{\ttdlocal}$ is recorded, namely for each edge of $\ttdtranssexpand$
that is adjacent to local state $\ttdlocal$, but only if it is real, or it
is an expansion edge that starts in local state $\ttdlocal$. Note that the
three \plkeyword{if} clauses in \algorithmref{Symbolically executing a path
  for local state l} are not disjoint: the first two both apply when edge
$e_i$ is ``vertical'': it both enters and exits local state~$\ttdlocal$. In
this case the two contributions cancel out.
\begin{algorithm}[t]
  \begin{algorithmic}[1]
    \REQUIRE path $\quotient \sigma = \range[]{t_1}{t_k}$ in $\ttdquotient$, i.e. $(t_i,t_{i+1}) \in \ttdtranssexpand$ for $1 \atm i < k$ ; local state $\ttdlocal$
    \STMT \code{$e_i$ := $(t_i,t_{i+1})$ for $1 \atm i < k$ , $(s_i,l_i)$ := $t_i$ for $1 \atm i \atm k$}
    \STMT \code{summary := "$n_{\ttdlocal}$"} \COMMENT{\code{summary} is a string}
    \FOR {\code{$i$ := $k-1$} \plkeyword{downto} 1}
      \IF {$e_i \in \ttdtranss$ and $l_i = \ttdlocal$}
        \STMT \code{summary := summary."+1"} \COMMENT{\code{.} = string concatenation}
      \ENDIF
      \IF {$e_i \in \ttdtranss$ and $l_{i+1} = \ttdlocal$}
        \STMT \code{summary := summary."-1"}
      \ENDIF
      \IF {$e_i \in \ttdtranssexpand \setminus \ttdtranss$ and $l_i = \ttdlocal$}
        \STMT \code{summary := summary."$\maxminus$1+1"}
      \ENDIF
    \ENDFOR
    \RETURN \code{summary}
  \end{algorithmic}
  \caption{Summary of a loop-free path segment}
  \label{algorithm: Symbolically executing a path for local state l}
\end{algorithm}

The summary of path $\quotient \sigma$ for local state $\ttdlocal$ defines
a function $\func{\Sigma_{\ttdlocal}}{\NN}{\NN}$ that summarizes the effect
of path $\quotient \sigma$ on counter $n_{\ttdlocal}$. The summary
functions for the short path in \figureref{vertical edge ttd} are shown
next to the figure. These examples illustrate how we can encode a loop-free
quotient path into a quantifier-free Presburger formula. The formula for
$\Sigma_0(n_0)$ implies that if we traverse the path backward from a state
with $n_0=0$ threads in local state 0, at the end there will be
$\Sigma_0(0) = 0 \maxminus 1 + 1 = 1$ thread in local state 0. If we start
with $n_0 = 1$, we also end up with $n_0 = 1$. Note that the path cannot be
traversed backward starting with $n_2 = 0$, since its endpoint is thread
state $\threadstate 2 2$.

Non-trivial SCC nodes along $\quotient \sigma$ are contractions of loops in
the expanded structure $\ttdexpand$, to the effect that paths in
$\ttdexpand$ are no longer finite; their summaries cannot be obtained by
symbolic execution. Instead we will determine a precise summary of simple
loops that is parameterized by the number $\iterator$ of times the loop is
executed. \Nonsimple\ loops are discussed in \sectionref{Pathwise
  Unbounded-Thread Reachability}.
\begin{figure}[t]
  \centerTwoOut{%
    \begin{minipage}{.4\textwidth}
      \begin{tikzpicture}[>=stealth, scale=1.0]
        \tikzstyle{vertex}=[circle,fill=white!25,minimum size=10pt,inner sep=-1pt]
        \draw[->,xshift=0cm] (0,0) -- coordinate (x axis mid) (3.2,0)node[anchor=north]{$l$};
        \draw[->,yshift=0cm] (0,0) -- coordinate (y axis mid) (0,2.7)node[anchor=east]{$s$};
        \foreach \x in {0,...,2}
        \foreach \y in {0,...,2}
        \node[vertex] (\y \x) at (\x*1.2+0.4,\y+0.3) {\scriptsize (\y,\x)};
        \draw[->, thick] (00) -- (10);
        \draw[->,dashed, thick] (10) -- (11);
        \draw[->, thick] (11) -- (22); 
      \end{tikzpicture}

    \end{minipage}}{%
    \begin{minipage}{.57\textwidth}
      Summary functions for local states $\ttdlocal=0,1,2$:
      \[
      \begin{array}{rclcl}
        \Sigma_0(n_0) & = & n_0 \maxminus 1 + 1 - 1 + 1 & = & n_0 \maxminus 1 + 1 \\
        \Sigma_1(n_1) & = & n_1 + 1 \\
        \Sigma_2(n_2) & = & n_2 - 1
      \end{array}
      \]
      Examples:
      \[
        \Sigma_0(0) = 1, \ \Sigma_0(1) = 1, \ \Sigma_1(0) = 1, \ \Sigma_2(1) = 0 \ .
      \]
    \end{minipage}}
  \caption{A loop-free quotient structure $\ttdquotient$ with a vertical real edge}
  \label{figure: vertical edge ttd}
\end{figure}

\section{Presburger Summaries for Simple Loops}
\label{section: Presburger Summaries for Simple Loops}

In this section we generalize path summaries to the case of simple SCCs,
formed by a single \emph{simple loop}, i.e., a single cyclic path without
repeated inner nodes. We aim at an exact solution in the form of a closed
expression for the value of local state counter $n_{\ttdlocal}$ after
\algorithmref{Abdulla} traverses the loop some number of times $\iterator$.

In this section, since we need to ``zoom in'' to SCCs collapsed into single
nodes in $\ttdquotient$, we instead look at paths in $\ttdexpand$. Recall
that for a loop-free path $\expand \sigma$, the value of counter
$n_{\ttdlocal}$ after \algorithmref{Abdulla} traverses $\expand \sigma$ can
be computed using $\expand \sigma$'s path summary function
$\Sigma_{\ttdlocal}$, determined via symbolic execution
(\algorithmref{Symbolically executing a path for local state l}). In the
case that $\expand \sigma$ is a loop, we would like to obtain a summary
formula parameterized by the number $\iterator$ of times the loop is
executed (we cannot replicate $\expand \sigma$'s summary function
$\iterator$ times, since $\iterator$ is a variable).

To this end, let $\expand \sigma = \range[]{t_1}{t_k}$ with $t_k = t_1$ be
a loop in $\ttdexpand$, and define $(s_i,l_i) := t_i$ for $1 \atm i \atm
k$. Let
\begin{equation}
  \begin{array}{rclc}
    \delta_{\ttdlocal} & \ = \ & |\{i: 1 \atm i < k: \ (t_i,t_{i+1}) \in \ttdtranss \ \land \ \pos[l]{$l_i$}{$l_{i+1}$} = \ttdlocal\}| & \ - \\
                       &       & |\{i: 1 \atm i < k: \ (t_i,t_{i+1}) \in \ttdtranss \ \land \                 l_{i+1}   = \ttdlocal\}| &
  \end{array}
  \label{equation: delta}
\end{equation}
be the \emph{real-edge summary} $\delta_{\ttdlocal} \in \ZZ$ of $\expand
\sigma$, i.e.\ the number of \emph{real} edges along $\expand \sigma$ that
start in local state $\ttdlocal$, minus the number of \emph{real} edges
along $\expand \sigma$ that end in~$\ttdlocal$. Value $\delta_{\ttdlocal}$
summarizes the total contribution by real edges to counter $n_{\ttdlocal}$
as path $\expand \sigma$ is traversed backward: real edges starting in
$\ttdlocal$ increment the counter, those ending in $\ttdlocal$ decrement
it. Let further $b_{\ttdlocal} = \Sigma_{\ttdlocal}(1)$ if $\expand \sigma$
ends in local state $\ttdlocal$ (in this case the backward traversal must
start with at least 1 thread in $\ttdlocal$), and $b_{\ttdlocal} =
\Sigma_{\ttdlocal}(0)$ otherwise.
\newcounterset{theoremCounterValues}{\theDEF}
\begin{THE}
  \label{theorem: counter values}
  Let superscript $\iterate{}$ denote $\iterator$ function applications.
  Then, for $\iterator \atl 1$,
  \begin{equation}
    \label{equation: counter values}
    \iterate{\Sigma_{\ttdlocal}}(n_{\ttdlocal}) = n_{\ttdlocal} \maxplus[b_{\ttdlocal}] \delta_{\ttdlocal} \maxplus[b_{\ttdlocal}] (\iterator-1) \cdot \delta_{\ttdlocal} \ .
  \end{equation}
\end{THE}
Recall that $\maxplus$ is not associative (it associates from left to
right); the right-hand side of \equationref{counter values} can generally
not be simplified to $n_{\ttdlocal} \maxplus[b_{\ttdlocal}] \iterator \cdot
\delta_{\ttdlocal}$.
Intuitively, the term $n_{\ttdlocal} \maxplus[b_{\ttdlocal}]
\delta_{\ttdlocal}$ marks the contribution to counter $n_{\ttdlocal}$ of
the first loop traversal, while $(\iterator-1) \cdot \delta_{\ttdlocal}$
marks the contribution of the remaining $\iterator-1$ traversals.

\paragraph{Example.}
We show how the \emph{un}reachability of thread state $\threadstate 6 4$
for the TTD in \figureref{ttd abstractions} is established.
For each local state $\ttdlocal \in \range 0 4$, the following constraints
are obtained (after simplifications) from summaries of the loop-free path
segment from $\threadstate 6 4$ to $\threadstate 3 1$ (``loop exit
point''), the loop inside the SCC node (black disc) using
\theoremref{counter values}, and the loop-free path segment from
$\threadstate 1 0$ (``loop entry point'') via $\threadstate 1 1$ to the
initial thread state $\threadstate 0 0$. Parameter $\iterator$ is the
number of times the loop is executed:
\[
\begin{array}{lrrrrrrrrrrrcr}
  \makebox[10mm][l]{$n_0:$} & 0 & \maxplus[0] &  0 & \maxplus[2] &  2 & \maxplus[2] & (\iterator-1) & \cdot &  2 & \maxplus[3] &  3 & \mbox{} \quad \atl \quad \mbox{} & 1 \\
                     n_1:   & 0 & \maxplus[1] &  0 & \maxplus[1] & -1 & \maxplus[1] & (\iterator-1) & \cdot & -1 & \maxplus[0] & -3 &         =                        & 0 \\
                     n_2:   & 0 & \maxplus[2] &  2 & \maxplus[0] & -1 & \maxplus[0] & (\iterator-1) & \cdot & -1 & \maxplus[0] &  0 &         =                        & 0 \\
                     n_3:   & 0 & \maxplus[0] & -2 & \maxplus[0] &  0 & \maxplus[0] & (\iterator-1) & \cdot &  0 & \maxplus[0] &  0 &         =                        & 0 \\
                     n_4:   & 1 & \maxplus[1] &  0 & \maxplus[0] &  0 & \maxplus[0] & (\iterator-1) & \cdot &  0 & \maxplus[0] &  0 &         =                        & 0
\end{array}
\]
The equation for $n_4$ simplifies to $1 = 0$ and thus immediately yields
unsatisfiability. Since there is only one path in $\ttdquotient$, we
conclude unreachability of $\ttdfinalstate = \threadstate 6 4$. In
contrast, for target thread state $\threadstate 6 3$, the equations for
$n_3$ and $n_4$ both reduce to $\true$. The conjunction of all five
equations reduces to $1 \maxplus[0] (\iterator-1) \cdot (-1) = 0$. This
formula is satisfied by $\iterator=2$, witnessing reachability of
$\threadstate 6 3$ via a path containing two full iterations of the loop
inside the SCC.

\section{Pathwise Unbounded-Thread Reachability}
\label{section: Pathwise Unbounded-Thread Reachability}

Consider an SCC along quotient path $\quotient \sigma$ that represents
several distinct simple loops in $\ttdexpand$. An example is an SCC with
two loops $A$ and $B$ that have one point in common and form an ``eight''
{\large $\infty$}. Such a double loop features paths of the form $(A|B)^*$,
where in each iteration there is a choice between $A$ and $B$. Our loop
acceleration technique from \sectionref{Presburger Summaries for Simple
  Loops} does not apply to such paths.

To solve this problem, we exploit the synergy between the pathwise analysis
suggested by the acyclic structure $\ttdquotient$, and the fact that
certain --- namely, simple --- paths can be processed using the technique
described in \sectionref{Presburger Summaries for Loop-Free Path Segments}
and~\sectionref[]{Presburger Summaries for Simple Loops}. \Nonsimple\ paths
are explored by \algorithmref{Abdulla}, but restricted to the narrow
``slice'' of $\ttd$ marked by the quotient path in $\ttdquotient$.

This algorithm is shown in \algorithmref{main}. It takes as input the TTD
$\ttd$, as well as the initial and final thread states, $\ttdinitstate$ and
$\ttdfinalstate$. The algorithm begins by building the quotient structure
$\ttdquotient$. This acyclic structure is now analyzed pathwise. For each
path $\quotient \sigma$ from $\ttdinitstate$ to $\ttdfinalstate$ in
$\ttdquotient$, we first decide whether it is \nonsimple\ or simple.
\begin{algorithm}[htbp]
  \begin{algorithmic}[1]
    \REQUIRE TTD $\ttd$, thread states $\ttdinitstate,\ttdfinalstate$
    \STMT $\ttdexpand$ \assign\ expanded TTD, $\ttdquotient$ \assign\ SCC quotient graph of $\ttdexpand$
    \FOREACH {path $\quotient \sigma$ in $\ttdquotient$ from $\ttdinitstate$ to $\ttdfinalstate$} \label{line: for each quotient path}
      \IF {$\quotient \sigma$ is \nonsimple}
        \IF {$\BWS(\familyobject{\infty}{(\restr{\ttd}{\quotient \sigma})},\union_{n \in \NN} \{\ttdinitshared\} \times \{\ttdinitlocal\}^n,\ttdfinalstate) = \mbox{``$\ttdfinalstate$ coverable''}$} \label{line: bws}
          \RETURN ``$\ttdfinalstate$ reachable in $\ttdinf$ from $\ttdinitstate$'' \label{line: reachable (1)}
        \ENDIF
      \ELSE
        \STMT $m$ \assign\ number of non-trivial SCCs visited by $\quotient \sigma$ \COMMENT{these SCCs are all simple}
        \STMT $\phi(\range[]{\iterator_1}{\iterator_m}) \ \assign \ \mbox{\em Presburger summary for $\quotient \sigma$}$ \COMMENT{Sect.~\sectionref[]{Presburger Summaries for Loop-Free Path Segments},~\sectionref[]{Presburger Summaries for Simple Loops}}
        \IF {$\phi(\range[]{\iterator_1}{\iterator_m})$ satisfiable} \label{line: Presburger encoding satisfiable}
          \RETURN ``$\ttdfinalstate$ reachable in $\ttdinf$ from $\ttdinitstate$'' \label{line: reachable (2)}
        \ENDIF
      \ENDIF
    \ENDFOR
    \RETURN ``$\ttdfinalstate$ unreachable in $\ttdinf$ from $\ttdinitstate$'' \label{line: unreachable}
  \end{algorithmic}
  \caption{Pathwise Reachability}
  \label{algorithm: main}
\end{algorithm}

\begin{enumerate}[$\bullet$]

\item If $\quotient \sigma$ is \nonsimple\ (visits some \nonsimple\ SCCs),
  we explore it using \algorithmref{Abdulla} (\lineref{bws}). More
  precisely, let $\restr{\ttd}{\quotient \sigma}$ be the restriction of the
  given TTD to the edges along $\quotient \sigma$, including any edges
  collapsed inside SCCs. Let further
  $\familyobject{\infty}{(\restr{\ttd}{\quotient \sigma})}$ be the
  infinite-state transition system derived from $\restr{\ttd}{\quotient
    \sigma}$ as described in \sectionref{Thread-Transition Diagrams and
    Backward Search}. We pass this transition system to procedure $\BWS$
  (\algorithmref{Abdulla}), along with the unchanged set of initial states,
  and the unchanged final state (which is also the end-point of $\quotient
  \sigma$). If this invocation results in ``coverable'', $\ttdfinalstate$
  is reachable in $\ttdinf$ from $\ttdinitstate$, which is hence returned
  in \lineref{reachable (1)}.

\item If $\quotient \sigma$ is simple (does not visit any
  \nonsimple\ SCCs), we can accelerate exploration along it using the
  techniques introduced in \sectionref{Presburger Summaries for Loop-Free
    Path Segments} and~\sectionref[]{Presburger Summaries for Simple
    Loops}. We build a Presburger summary for the path, parameterized by
  the loop iteration counts $\iterator_i$, one for each
  loop.\footnote{Loop-free paths ($m=0$) can be processed either using
    \algorithmref{Abdulla}, or via summaries.} If this formula is
  satisfiable, again we have that $\ttdfinalstate$ is reachable in
  $\ttdinf$ from $\ttdinitstate$. The assignment to the $\iterator_i$ gives
  the number of times each loop needs to be traversed; from this data a
  multi-threaded path through $\ttd$ can easily be constructed.
\end{enumerate}
If none of the paths $\quotient \sigma$ results in the answer ``coverable''
by either concrete or symbolic exploration, $\ttdfinalstate$ is unreachable
in $\ttdinf$ from $\ttdinitstate$, which is hence returned as the answer.
Note that this happens in particular if there is no path at all from
$\ttdinitstate$ to $\ttdfinalstate$ in $\ttdquotient$.

\paragraph{Correctness.} \algorithmref{main} terminates since
$\ttdquotient$ is acyclic, so the loop in \lineref{for each quotient path}
goes through finitely many iterations. Partial correctness follows from the
following two claims. Let $\quotient \sigma$ be a quotient path considered
in \lineref{for each quotient path}.
\begin{enumerate}[1.]

\item \label{claim: nonsimple}
  If $\quotient \sigma$ is \nonsimple, then \algorithmref{main} outputs
  ``reachable'' in \lineref{reachable (1)} iff $\ttdfinalstate$ is
  reachable in $\ttdinf$ along the edges represented by $\quotient \sigma$.

\item \label{claim: simple}
  If $\quotient \sigma$ is simple, then \algorithmref{main} outputs
  ``reachable'' in \lineref{reachable (2)} iff $\ttdfinalstate$ is
  reachable in $\ttdinf$ along the edges represented by $\quotient \sigma$.

\end{enumerate}
Claim~\ref{claim: nonsimple} is proved using soundness and completeness of
\algorithmref{Abdulla}. Claim~\ref{claim: simple} is proved using
\theoremref{counter values}.
Given these claims, we obtain:
\begin{COR}[Soundness]
  \label{corollary: soundness}
  If \algorithmref{main} returns ``reachable'' (\lineref{reachable (1)}
  or~\lineref[]{reachable (2)}) or ``unreachable'' (\lineref{unreachable}),
  then $\ttdfinalstate$ is reachable or unreachable, respectively, in
  $\ttdinf$.
\end{COR}
\proof \mbox{}
\begin{enumerate}[$\bullet$]

\item  If \algorithmref{main} returns ``reachable'', then it does so for
some $\quotient \sigma$, in Lines~\lineref[]{reachable (1)}
or~\lineref[]{reachable (2)}. The fact that $\ttdfinalstate$ is actually
reachable in $\ttdinf$ follows from one of the two claims above, depending
on whether $\quotient \sigma$ is \nonsimple\ or simple.

\item If \algorithmref{main} returns ``unreachable'', then it does not
reach Lines~\lineref[]{reachable (1)} or~\lineref[]{reachable (2)}, for
\emph{any}~$\quotient \sigma$. By the above two claims, $\ttdfinalstate$ is
not reachable in $\ttdinf$ along the edges represented by any quotient
path. The fact that then $\quotient \sigma$ is not reachable in $\ttdinf$
at all follows from the proof of \lemmaref{quotient structure
  overapproximates}: the proof shows that, if $\ttdfinalstate$ is
reachable, then there exists a quotient path in $\ttdquotient$ from
$\ttdinitstate$ to $\ttdfinalstate$ such that $\ttdfinalstate$ is reachable
in $\ttdinf$ along the edges represented by that quotient
path.\eopBox
\end{enumerate}
\paragraph{Implementation.} Our technique is implemented in a reachability
checker named \ourtool\footnote{\ourtool\ ``='' \textbf{C}oncolic
  \textbf{U}nbounded-\textbf{T}hread \textbf{R}eachability
  analysis.}. We discuss some details on the
implementation of \algorithmref{main} in \ourtool.

\lineref{for each quotient path} selects potential paths in $\ttdquotient$.
Since we can abort the algorithm once a path is found that witnesses
reachability, it makes sense to rank the paths by ``promise'' of ease of
processing: we begin with loop-free paths, i.e.\ those with only trivial
SCCs, followed by paths with simple SCCs whose edges are all real, followed
by paths with simple SCCs that feature expansion edges. Finally we select
paths with \nonsimple\ loops inside SCCs. The length of a path is
secondary.

In order to call $\BWS$ in \lineref{bws} on the TTD restricted to the edges
represented by $\quotient \sigma$, there is no need to construct
$\restr{\ttd}{\quotient \sigma}$ a priori. Instead, when computing cover
preimages, we make sure to only fire TTD edges belong $\quotient \sigma$
and its loops.

\ifspace{To keep our computational model simple, we have excluded from the
  formalization in \sectionref{Thread-Transition Diagrams and Backward
    Search} \emph{dynamic thread creation}, where threads are spawned
  during the execution of the program. This feature does not formally add
  expressive power, but is often included for its presence in
  multi-threaded software. Our implementation does support thread creation.
  Symbolically backward-executing a thread creation edge is
  straightforward: the counter of the local state of the spawned thread
  must be decreased, since that thread does not exist in the source state.
  Our implementation performs some book-keeping to ensure the
  backward-executability of such an edge: both the local state of the
  spawned thread, as well as that of the spawning thread must exist in the
  successor state, since the spawning thread does not change its state (it
  only side-effects the thread creation).}

\section{Empirical Evaluation}
\label{section: Empirical Evaluation}

In this section we provide experimental results obtained using \ourtool.
The goal of the experiments is to measure the performance impact of the
presented approach compared to the backward search \algorithmref{Abdulla}.
We expect our approach to improve the latter,
as it is short-cutting standard backward exploration across simple loops
and linear path segments. The question is whether solving Presburger
equations instead of concretely exploring loops actually amounts to speed-up.

\subsubsection{Experimental Setup.}

We collected an extensive set of benchmarks, 173 in total, which is
organized into two suites. The first suite contains 124 TTDs obtained from
Boolean programs (BPs), which are in turn obtained from \C\ source programs
(taken from~\cite{KKW14a}) via predicate abstraction. As TTDs are
equivalent in expressiveness to certain forms of Petri
nets~\cite{ELMMN14,KKW14a}, we include PN examples in our
\noindent
\centerTwoOut{%
  \begin{minipage}{93mm}
    benchmark collection. The second suite therefore contains 49 TTDs
    obtained from PNs (taken from~\cite{ELMMN14}). While PNs are not the
    main focus of this work, we were hoping to get insights into how
    complex concurrency affects our approach, as the PNs available to us
    exhibited more challenging concurrent behaviors than the BPs.
    The table on the right shows size ranges of the benchmarks.
  \end{minipage}}{%
  \begin{minipage}{27mm}
    \resizebox{2.7cm}{!}{
      \raggedright
      \begin{tabular}{|c||r|r|}
        \hline
        \multicolumn{1}{|c||}{~BP~} & ~min.~ & max.~ \\
        \hline
        $|\ttdshareds|$   &  $5$   & $257$ \\
        $|\ttdlocals| $   &  $14$  & $4097$ \\
        $|\ttdtranss| $   &  $18$  & ~$20608$ \\ 
        \hline
        \hline
        ~PN~ & ~min.~ & max.~ \\
        \hline
        $|\ttdshareds|$   &  $6$   & $18234$ \\
        $|\ttdlocals| $   &  $6$   & $332$ \\
        $|\ttdtranss| $   &  $13$  & ~$27724$ \\ 
        \hline
      \end{tabular}}
  \end{minipage}}
 
\vskip2pt 
We use Z3 (v4.3.2)~\cite{MB08} as the Presburger solver. For
each benchmark, we consider verification of a safety
property. In the case of BP examples, the property is specified via an
assertion. There are 87 safe instances in total: 56 of the BPs, and 31 of
the PNs. All
experiments are performed on a 2.3GHz Intel Xeon machine with 64 GB memory,
running 64-bit Linux. Execution time is limited to 30min, and memory to 4
GB. All benchmarks and our tool are available
online\footnote{Webpage: \ourtoolurl}. 

\begin{figure}[t]
  \centering
  \pgfplotsset{every tick label/.append style={font=\footnotesize},
  label style={font=\footnotesize}}
\pgfplotsset{compat=1.3} 
\Ifthen{\fillinbtw}{\usepgfplotslibrary{fillbetween}}
\begin{minipage}{.49\linewidth}
  \begin{tikzpicture}[scale=0.76]
    \begin{loglogaxis}[
      height=7.6cm,
      width= 7.6cm,
      legend pos=north west,
      legend style={font=\scriptsize},        
      xmin=0.01,xmax=3600,
      ymin=0.01,ymax=3600,
      xticklabels={,,,$10^{0}$,$10^{1}$,$10^{2}$,$10^{3}$,TO},
      xlabel={\scriptsize \BWS\ (sec.)},
      yticklabels={,,,$10^{0}$,$10^{1}$,$10^{2}$,$10^{3}$,TO},
      extra x ticks={3600},
      extra x tick style={
        tick label style={rotate=90} 
      },
      extra x tick label={\tiny RO},
      extra y ticks={3600},
      extra y tick label={\scriptsize RO},
      ylabel={\scriptsize \ourtool\ (sec.)}]
      \addplot[only marks, mark=x, color=blue] table {data/bp-unsafe-t.dat};
      \addplot[only marks, mark=o, color=red] table {data/bp-safe-t.dat};
      \draw[red] (axis cs:0.01,0.01) -- (axis cs:3600,3600);
      \draw[gray, dashed] (axis cs:0.01,1) -- (axis cs:1,1);
      \draw[gray, dashed] (axis cs:1,0.01) -- (axis cs:1,1);
      \legend{unsafe BP, ~~~safe BP}
    \end{loglogaxis}
  \end{tikzpicture}
\end{minipage}
\hfill{
  \begin{minipage}{.49\linewidth}
    \begin{tikzpicture}[scale=0.76]
      \begin{loglogaxis}[
        height=7.6cm,
        width= 7.6cm,
        legend pos=north west,
        legend style={font=\scriptsize},    
        xmin=0.01,xmax=3600,
        ymin=0.01,ymax=3600,
        xticklabels={,,,$10^{0}$,$10^{1}$,$10^{2}$,$10^{3}$,TO},
        xlabel={\scriptsize \BWS\ (sec.)},
        yticklabels={,,,$10^{0}$,$10^{1}$,$10^{2}$,$10^{3}$,TO},
        extra x ticks={3600},
        extra x tick style={
          tick label style={rotate=90} 
        },
        extra x tick label={\tiny RO},
        extra y ticks={3600},
        extra y tick label={\scriptsize RO},
        ylabel={\scriptsize \ourtool\ (sec.)}]
        \addplot[only marks, mark=x, color=blue] table {data/pn-unsafe-t.dat};
        \addplot[only marks, mark=o, color=red] table {data/pn-safe-t.dat};
        \draw[red] (axis cs:0.01,0.01) -- (axis cs:3600,3600);
        \draw[gray, dashed] (axis cs:0.01,1) -- (axis cs:1,1);
        \draw[gray, dashed] (axis cs:1,0.01) -- (axis cs:1,1);
        \legend{unsafe PN, ~~~safe PN}
      \end{loglogaxis}
    \end{tikzpicture}
  \end{minipage}}
\hfill{
  \begin{minipage}{0.49\linewidth}
    \begin{tikzpicture}[scale=0.76]
      \begin{semilogyaxis}[
        height=7.6cm,
        width= 7.6cm,
        legend columns=1,
        legend pos=north west, 
        legend style={font=\scriptsize},      
        every axis plot post/.append style={mark=none},
        xmin=1,xmax=124,
        xtick={20,40,60,80,100,124},
        xlabel={\scriptsize BP Benchmark},
        ymin=1,ymax=4096,
        extra y ticks={4096},
        extra y tick label={\scriptsize RO},
        yticklabels={,$10^{1}$,$10^{2}$,$10^{3}$},
        ylabel={\scriptsize Memory Usage (MB.)}] 
        \ifthenelse{\fillinbtw}{
          \addplot [name path=A, solid, color=  red] table[x index=0, y index=1] {data/bp-m.dat};
          \addplot [name path=B, solid, color=blue] table[x index=0, y index=2] {data/bp-m.dat};
          \addplot[red!10] fill between[of=A and B];
        }{
          \addplot [solid, color= red] table[x index=0, y index=1] {data/bp-m.dat};
          \addplot [solid, color=blue] table[x index=0, y index=2] {data/bp-m.dat};
        }
        \draw[gray, dashed] (axis cs:1,10) -- (axis cs:115,10);
        \legend{\BWS, \ourtool}
      \end{semilogyaxis}
    \end{tikzpicture}
  \end{minipage}}
\hfill{
  \begin{minipage}{0.49\linewidth}
    \begin{tikzpicture}[scale=0.76]
      \begin{semilogyaxis}[
        height=7.6cm,
        width= 7.8cm,
        legend columns=1,
        legend pos=north west, 
        legend style={font=\scriptsize},      
        every axis plot post/.append style={mark=none},
        xmin=1,xmax=49,
        xtick={10,20,30,40,49},
        xlabel={\scriptsize PN Benchmark},
        ymin=1,ymax=4096,
        extra y ticks={4096},
        extra y tick label={\scriptsize RO},
        yticklabels={,$10^{1}$,$10^{2}$,$10^{3}$},
        ylabel={\scriptsize Memory Usage (MB.)}]
        \ifthenelse{\fillinbtw}{
          \addplot [name path=A, solid, color=  red] table[x index=0, y index=1] {data/pn-m.dat};
          \addplot [name path=B, solid, color=blue] table[x index=0, y index=2] {data/pn-m.dat};
          \addplot[red!10] fill between[of=A and B];   
        }{
          \addplot [solid, color= red] table[x index=0, y index=1] {data/pn-m.dat};
          \addplot [solid, color=blue] table[x index=0, y index=2] {data/pn-m.dat};
        }
        \draw[gray, dashed] (axis cs:1,10) -- (axis cs:55,10);
        \legend{\BWS, \ourtool}
      \end{semilogyaxis}
    \end{tikzpicture}
  \end{minipage}}

  \caption[caption]{Performance impact (BP/PN = TTD from BP/PN). RO stands
    for ``out of resources'': the run reached the time or the memory limit
    before producing a result.
    
    \setlength{\parindent}{2em}

    $\bullet$ Top row shows the comparisons of execution time. Left:
    comparison on BPs; Right: comparison on PNs. Each plot represents
    execution time on one example.

    $\bullet$ Bottom row shows the comparison of memory usage. Left:
    comparison on BPs; Right: comparison on PNs. The curves are sorted by
    the memory usage of \ourtool.}
  \label{figure: bws vs cutr}
\end{figure}

\subsubsection{Comparison.} We first consider TTDs obtained from BPs,
the target of this work. The runtime comparison results are given
in the top left part of \figureref{bws vs cutr}. The results demonstrate
that \ourtool\ performs much better than \BWS. In some cases, runs that
time out in \BWS\ can be successfully solved by \ourtool\ within 30min; in
contrast, there is only one example such that \BWS\ successfully completes
while \ourtool\ runs time out. The latter situation can be explained by the
path explosion: there are more than 5000 paths for this example.

We now consider TTDs obtained from PNs; see top right of \figureref{bws vs
  cutr}. Here we see little performance difference. Investigating this
further, we found that for Petri nets the density of TTD edges is higher
(explained by their more complex structure and the relatively less
organization and control in the concurrent systems represented by Petri
nets, compared to programs). This has two consequences: (i) there are few
but large SCCs, and (ii) most of them are \nonsimple. As a result, there
are few paths through the quotient structure, and almost all of them are
explored via calling $\BWS$. This makes the whole process essentially
equivalent to a single call to $\BWS$.


The curves at the bottom left of \figureref{bws vs cutr} illustrate that
\BWS\ utilizes less memory on small BP benchmarks, an effect can be
explained by the overhead of pathwise analysis and Z3. On large examples
\ourtool\ tends to need less memory resource than \BWS. The memory
comparison for PNs shows similar results.

The performance impact of our acceleration approach, on both runtime and
memory, can be summarized as follows. Our method analyzes a specific path
at a time. If $\ttdfinalstate$ is reachable, there is a good chance
\ourtool\ can find a solution early, due to the ranking of paths, some of
which permit quick decisions.
%
%
Although \ourtool\ relies on backward search to cross nested loops, the
cost of that is limited as such exploration is confined to a small fragment
of the TTD. In the extreme, the entire TTD contains only one path with
\nonsimple\ loops. In this case \ourtool\ falls back on backward search.

\section{Related Work}

Groundbreaking results in
infinite-state system analysis include the
decidability of coverability in \emph{vector addition systems} (VAS)
\cite{KM69}, and the work by German and Sistla on modeling
communicating finite-state threads as VAS~\cite{GS92}. Numerous
results have since
improved on the original
procedure in \cite{KM69} in practice \cite{GRB06,GRB07,RS11,VH12}.
Others extend it to more general computational models, including
\emph{well-structured} \cite{FS01} or \emph{well quasi-ordered} 
transition systems \cite{ACJT96,A10}.

Recent theoretical work by Leroux employs Presburger arithmetic to solve
the VAS global configuration reachability (not coverability) problem. In
\cite{L09}, it is shown that a state is \emph{unreachable} in the
VAS iff there exists an ``inductive'' Presburger formula that separates the
initial and final states.
The theoretical complexity of this technique is mostly left open.
Practicality is not discussed and doubted later by the author in
\cite{L13}, where a more direct approach is presented that permits
the computation of a Presburger definition of the reachability set of the
VAS in some cases, e.g.\ for \emph{flatable} VAS. Reachability can then be
cast as a Presburger decision problem.
The question under what exact conditions the VAS reachability set is
Presburger-definable appears to be undecided.

The results referenced above are mainly foundational in nature and target
generally harder (even undecidable) reachability questions than we do in
this paper. We emphasize that our motivation for acceleration is not to
ensure convergence of (otherwise possible diverging) fixpoint computations.
Instead, our goal here was to show, for the decidable problem of TTS thread
state reachability, (i) how to practically compute a Presburger encoding
whose satisfiability implies reachability of the thread state, and (ii)
that the resulting (quantifier-free) formulas are often
easy to decide, thus giving rise to an efficient algorithm. Existing
(typically forward) acceleration techniques for infinite-state systems
\mbox{\cite{FL02,BFLS05,JS07}} were inspirational for this paper.

In recent work, Petri net \emph{marking equations}
are used to reduce the coverability problem to linear
constraint solving~\cite{ELMMN14}. Follow-up work investigates a similar
approach for thread-transition systems \cite{ALW16}.
Like the present work, these approaches
benefit from advances in SMT technology and in fact have proved to be
efficient. On the other hand, they are incomplete (the constraints
overapproximate coverability). Our goal here was
to retain (soundness and) completeness, and to investigate at what cost
this can be achieved.

\section{Conclusion}
\label{section: Conclusion}

In this paper, we have presented an approach for accelerating a
widely-applicable
infinite-state search algorithm for systems of unbounded numbers of
threads. A~key ingredient
is the construction of an acyclic quotient of the input program, which in
turn enables a finite path-by-path analysis.
Loop-free paths and paths with only simple loops can be collapsed
\emph{without approximation} into Presburger arithmetic constraints that
symbolically summarize the effect of executing the backward search
algorithm along these paths in the multi-threaded program. Each path
passing through loop nests is processed via standard explicit-state
backward search but confined to this particular path. We have demonstrated
the power of this method for proving and refuting safety properties of an
extensive set of TTDs obtained from Boolean program benchmarks.
We conclude that partial but exact symbolic acceleration of existing sound
and complete infinite-state search algorithms is very much feasible, and in
fact very beneficial.


\bibliographystyle{splncs03}
\bibliography{paper}

\Ifthen{\arxivVersion}{

\newpage

\appendix

\section*{Appendix}

\section{Uniqueness of the Initial State}
\label{appendix: Uniqueness of the Initial State}

In our problem formulation we have assumed a unique initial state of the
program (TTD) executed by each thread. This essentially fixes the start
state of each TTD path we have to process. The more general case of
multiple initial states can of course be handled by considering each
initial state in turn. However, in many practical cases, the general case
can be reduced to the unique-initial-thread-state case without affecting
thread state reachability. Suppose the initial thread state set $T$ of the
given TTD $\ttd$ satisfies the following ``box'' property:
\begin{equation}
  \forall (s,t) \in T, \ (s',t') \in T: \ (s,t') \in T, \ (s',t) \in T \ .
  \label{equation: box property}
\end{equation}
This holds if $T$ is (already) a singleton. More generally, it holds if all
states in $T$ have the same shared state,\footnote{Shared states are formed
  by valuations of shared (global) Boolean variables, which in turn store
  values of predicates over \C\ program variables. If these \C\ program
  variables are themselves global, they are initialized by the operating
  system and hence unique.} and it holds if all states in $T$ have the same
local state. It also holds of a set $T$ whose elements form a complete
rectangle in the graphical representation of $\ttd$.

To enforce a unique initial thread state, we build a new TTD $\ttd'$ that
is identical to $\ttd$, except that it has a single initial thread state
$\ttdinitstate = \ttdinitstatepair$ with fresh shared and local states
$\ttdinitshared,\ttdinitlocal$, and the following additional edges:
\begin{eqnarray}
  \ttdinitstatepair            \rightarrow \threadstate s l & \quad \mbox{such that $\threadstate s l \in T$ ,} & \quad \mbox{and} \label{equation: additional edges 1} \\
  \threadstate s \ttdinitlocal \rightarrow \threadstate s l & \quad \mbox{such that $\threadstate s l \in T$ .}                    \label{equation: additional edges 2}
\end{eqnarray}
We see that the cost of obtaining $\ttd'$ from $\ttd$ is linear in the size
of the initial state set $T$.

Suppose now some thread state $t_0 = \threadstate{s_0}{l_0}$ is reachable
in $\ttd[n]$, for some~$n$. Then there exists a path from some global state
$\state{\ttdshared[J]}{\range[]{l_1}{l_n}}$ such that $\threadstate
{\ttdshared[J]} {l_i} \in T$ for all $i$, to a global state with shared
component $s_0$ and some thread in local state~$l_0$. We can attach, to the
front of this path, the prefix
\begin{eqnarray*}
  \state{\ttdinitshared}{\range[]{\ttdinitlocal}{\ttdinitlocal}} & \wbox{$\transsymbol$} & \state{\underline{\ttdshared[J]}}{\underline{l_1},\ttdinitlocal,\range[]{\ttdinitlocal}{\ttdinitlocal}} \\
                                                        & \transsymbol & \state{           \ttdshared[J]} {l_1,\underline{l_2},\range[]{\ttdinitlocal}{\ttdinitlocal}} \\
                                                        &              & \quad \quad \cdots \\
                                                        & \transsymbol & \state{           \ttdshared[J]} {l_1,l_2,\range[]{l_3}{\underline{l_n}}} \ ,
\end{eqnarray*}
with the underlined symbols changed. The new path reaches $t_0$ in
$\ttd[n]'$.

Conversely, suppose some thread state $t_0 = \threadstate{s_0}{l_0}$ such
that $s_0 \not= \ttdinitshared$, $l_0 \not= \ttdinitlocal$ is reachable in
$\ttd[n]'$, for some~$n$. Then there exists a path $p'$ from
$\{\ttdinitshared\} \times \{\ttdinitlocal\}^n$ to a global state with shared
component $s_0$ and some thread in local state~$l_0$. The very first
transition of $p'$ is by some thread executing an edge of type
\equationref[]{additional edges 1}, since those are the only edges leaving
the unique initial state $\ttdinitstatepair$. Let that be thread number
$i$, and let $\threadstate s l \in T$ be the new state of thread $i$.

Consider now an arbitrary thread $j \in \{1,\ldots,n\} \setminus \{i\}$;
its local state after the first transition along $p'$ is $\ttdinitlocal$.
\begin{enumerate}[$\bullet$]

\item If thread $j$ is never executed along $p'$, we build a new path $p''$
  by inserting edge $\edge s \ttdinitlocal s l$, executed by thread $j$, right
  after the first transition in~$p'$. This is a valid edge (of type
  \equationref[]{additional edges 2}) since $\threadstate s l \in T$. The
  edge moves thread $j$ into an initial thread state $(s,l) \in T$. The
  modified state sequence remains a valid path in $\ttd[n]'$ since no
  shared states have been changed, and thread $j$ is inactive henceforth.

\item If thread $j$ is executed along $p'$, then the first edge it executes
  must be of type \equationref[]{additional edges 2}, since again this is
  the only way to get out of local state $\ttdinitlocal$. Let
  $\threadstate{\bar s}{\bar l} \in T$ be the state of thread $j$ after
  executing this first edge. Then $\edge s \ttdinitlocal s {\bar l}$ is a
  valid edge (of type \equationref[]{additional edges 2}): from
  $\threadstate s l \in T$ and $\threadstate{\bar s}{\bar l} \in T$, we
  conclude $\threadstate s {\bar l} \in T$, by property~\equationref[]{box
    property}. We now build a new path $p''$, by removing from $p'$ thread
  $j$'s first transition, and instead inserting, right behind the first
  transition of $p'$, a transition where thread $j$ executes edge $\edge s
  \ttdinitlocal s {\bar l}$:
  \[
  \begin{array}{cccl}
    & p'   & :: & \edge [i] \ttdinitshared \ttdinitlocal s t \ , \quad \ldots \quad , \edge [j] {\bar s} \ttdinitlocal {\bar s} {\bar l} \\[1mm]
    \mbox{becomes} \\
    & p''  & :: & \edge [i] \ttdinitshared \ttdinitlocal s t \ , \edge [j] s \ttdinitlocal s {\bar l} \ , \quad \ldots
  \end{array}
  \]
  (here we add a thread index on top of an edge's arrow, to indicate the
  identity of the executing thread). The modified state sequence remains a
  valid path in $\ttd[n]'$, since the shared states ``match'' and are not
  changed by any of the removed or inserted edges. Moving the local state
  change of thread $j$ (from $\ttdinitlocal$ to $\bar l$) forward leaves
  the path intact, since the original edge $\edge {\bar s} \ttdinitlocal
  {\bar s} {\bar l}$ was thread $j$'s first activity.

\end{enumerate}
This procedure is applied to every thread $j \not= i$, with the result
that, after the first $n$ transitions, all threads are in a state belonging
to $T$. The suffix of $p''$ following these transitions reaches $t_0$ in
$\ttd[n]$.\eop

\section{Proof of Lemma \lemmaref[]{quotient structure overapproximates}}
\label{appendix: Proof of Lemma quotient structure overapproximates}

\setcounter{DEF}{\thelemmaCounterValues}
\begin{LEM}
  \label{lemma: proof: quotient structure overapproximates}
  If thread state $\ttdfinalstate$ is reachable in $\ttdinf$, then
  $\ttdfinalstate$ is also reachable in $\ttdquotient$.
\end{LEM}
\Proof: We show that $\ttdfinalstate$ is reachable in $\ttdexpand$; the fact
that $\ttdfinalstate$ is reachable in $\ttdquotient$ then follows
from standard properties of the SCC quotient graph.

Let $\ttdfinalstate=\ttdfinalstatepair$, and
$\ttdinitstate=\ttdinitstatepair$ be the initial state. Since
$\ttdfinalstate$ is reachable in $\ttdinf = \union_{n=1}^\infty \ttd[n]$,
let $n$ be such that $\ttdfinalstate$ is reachable in $\ttd[n]$ via a
witness path $p$:
\begin{equation}
  \label{equation: witness path}
  p \ :: \ \state{\ttdinitshared}{\underbrace{\ttdinitlocal, \ldots, \ttdinitlocal}_n} \quad \transsymbol \quad \cdots \quad \transsymbol \quad \state \ttdfinalshared {l_1, \ldots, l_{i-1}, \ttdfinallocal, l_{i+1}, \ldots, l_n} .
\end{equation}
Let further $(e_i) := (e_1,\ldots,e_z)$ be the sequence of TTD edges
executed along~$p$. We drop all ``horizontal'' edges from $(e_i)$,
i.e.\ edges of the form $\edge s \cdot s \cdot$, to obtain a subsequence
$(g_i) := (g_1,\ldots,g_{z'})$ ($z' \atm z$). Given $(g_i)$, we construct a
path $\sigma$ from $\ttdinitstate$ to $\ttdfinalstate$ in $\ttdexpand$, by
\emph{processing} the edges $g_i$, defined recursively as follows:
\begin{enumerate}[(1)]

\item Edge $g_1$ is processed by copying it to $\sigma$.

\item Suppose edge $g_{k-1}$ has been processed, and suppose its target
  state is $(s,l_i)$. Edge $g_k$'s source state has shared component $s$ as
  well, since edges $g_{k-1}$ and $g_k$ are consecutive in $p$, except
  for some horizontal edges in between that may have been dropped, but
  these do not change the shared state. So let $g_k$'s source state be
  $(s,l_j)$.

  Edge $g_k$ is now processed as follows. If $l_i=l_j$, append $g_k$ to
  $\sigma$. Otherwise, first append $\expedge s {l_i} s {l_j}$ to $\sigma$,
  then $g_k$. Note that $\expedge s {l_i} s {l_j}$ is a valid expansion
  edge in $\expand R$, since there exist two non-horizontal edges,
  $g_{k-1}$ and $g_k$, adjacent to the expansion edge's source and target,
  respectively.
\end{enumerate}
Step (2) is repeated until all edges $g_i$ have been processed. It is clear
by construction that $\sigma$ is a valid path in $\ttdexpand$, and that it
starts in $\ttdinitstate = \ttdinitstatepair$. We finally have to show that
it ends in $\ttdfinalstate = \ttdfinalstatepair$. It may in fact not: let
$(\ttdfinalshared,l_f)$ be the target state of the final edge $g_{z'}$;
$l_f$ may or may not be equal to $\ttdfinallocal$. If it is not, we append
an edge $\expedge \ttdfinalshared {l_f} \ttdfinalshared \ttdfinallocal$ to
$\sigma$. This is a valid expansion edge by \definitionref{expanded ttd},
and $\sigma$ now ends in $\ttdfinalstate$, which is hence reachable in
$\ttdexpand$.\eop

\section{Proof of Theorem \theoremref[]{counter values}}
\label{appendix: Proof of Theorem counter values}

Before we turn to this proof, we establish a lemma that uses the
$\delta_{\ttdlocal}$'s defined in \sectionref{Presburger Summaries for
  Simple Loops} to compactly determine local state $\ttdlocal$'s
summary along $\expand \sigma$.
\begin{LEM}
  \label{lemma: counter values single iteration}
  Let $b_{\ttdlocal} = \Sigma_{\ttdlocal}(1)$ if $l_k = \ttdlocal$ (path
  $\expand \sigma$ ends in local state $\ttdlocal$), and $b_{\ttdlocal} =
  \Sigma_{\ttdlocal}(0)$ otherwise. Then $\Sigma_{\ttdlocal}(n_{\ttdlocal})
  = n_{\ttdlocal} \maxplus[b_{\ttdlocal}] \delta_{\ttdlocal}$ .
\end{LEM}
The lemma suggests: in order to determine local state $\ttdlocal$'s summary
function in compact form, first compute the constant
$\Sigma_{\ttdlocal}(1)$ (or $\Sigma_{\ttdlocal}(0)$) using
\algorithmref{Symbolically executing a path for local state l}.
$\Sigma_{\ttdlocal}(n_{\ttdlocal})$ is then the formula as specified in the
lemma.

\

\noindent
\Proof\ of \lemmaref{counter values single iteration}: by induction on the
number $k$ of vertices of $\expand \sigma = \range[]{t_1}{t_k}$.

\fbox{$k=1$:} then $\expand \sigma$ has no edges, so
$\Sigma_{\ttdlocal}(n_{\ttdlocal}) = n_{\ttdlocal}$, $b_{\ttdlocal} = 0$,
and $\delta_{\ttdlocal} = 0$. Thus, $\Sigma_{\ttdlocal}(n_{\ttdlocal}) =
n_{\ttdlocal} = n_{\ttdlocal} \maxplus[b_{\ttdlocal}] 0 = n_{\ttdlocal}
\maxplus[b_{\ttdlocal}] \delta_{\ttdlocal}$.

\fbox{$k \rightarrow k+1$:} Suppose $\expand \sigma =
\range[]{t_1}{t_{k+1}}$ has $k+1$ vertices, and \lemmaref{counter values
  single iteration} holds for all paths of $k$ vertices. One such path is
the suffix $\expand \tau = \range[]{t_2}{t_{k+1}}$ of $\expand \sigma$. By
the induction hypothesis, $\expand \tau$'s summary function
$\Tau_{\ttdlocal}$ satisfies $\Tau_{\ttdlocal}(n_{\ttdlocal}) =
n_{\ttdlocal} \maxplus[c_{\ttdlocal}] \gamma_{\ttdlocal}$ for the real edge
summary $\gamma_{\ttdlocal}$ along $\expand \tau$, and $c_{\ttdlocal} =
\Tau_{\ttdlocal}(1)$ if $l_{k+1} = \ttdlocal$; otherwise $c_{\ttdlocal} =
\Tau_{\ttdlocal}(0)$. Note that $\expand \tau$ and $\expand \sigma$ have
the same final state $t_{k+1} = (s_{k+1},l_{k+1})$.

We now distinguish what \algorithmref{Symbolically executing a path for
  local state l} does to the first edge $e_1 = (t_1,t_2) =
((s_1,l_1),(s_2,l_2))$ of $\expand \sigma$ (which is traversed last):
\begin{description}

\item[Case 1:] $e_1 \in \ttdtranss$ and $l_1 = \ttdlocal$: Then
  $\Sigma_{\ttdlocal}(n_{\ttdlocal}) = \Tau_{\ttdlocal}(n_{\ttdlocal}) +
  1$, $\delta_{\ttdlocal} = \gamma_{\ttdlocal} + 1$, and $b_{\ttdlocal} =
  c_{\ttdlocal} + 1$. Using the induction hypothesis (IH), we get
  $\Sigma_{\ttdlocal}(n_{\ttdlocal}) = n_{\ttdlocal}
  \maxplus[c_{\ttdlocal}](\delta_{\ttdlocal} - 1) + 1$.
  \begin{itemize}

  \item If $n_{\ttdlocal} + \delta_{\ttdlocal} - 1 \atl c_{\ttdlocal}$,
    then $n_{\ttdlocal} \maxplus[c_{\ttdlocal}](\delta_{\ttdlocal} - 1) + 1
    = n_{\ttdlocal} + \delta_{\ttdlocal} = n_{\ttdlocal}
    \maxplus[b_{\ttdlocal}] \delta_{\ttdlocal}$ since $n_{\ttdlocal} +
    \delta_{\ttdlocal} \atl c_{\ttdlocal} + 1 = b_{\ttdlocal}$.

  \item If $n_{\ttdlocal} + \delta_{\ttdlocal} - 1 < c_{\ttdlocal}$, then
    $n_{\ttdlocal} \maxplus[c_{\ttdlocal}](\delta_{\ttdlocal} - 1) + 1 =
    c_{\ttdlocal} + 1 = b_{\ttdlocal} = n_{\ttdlocal}
    \maxplus[b_{\ttdlocal}] \delta_{\ttdlocal}$ since $n_{\ttdlocal} +
    \delta_{\ttdlocal} < c_{\ttdlocal} + 1 = b_{\ttdlocal}$.

  \end{itemize}

\item[Case 2:] $e_1 \in \ttdtranss$ and $l_2 = \ttdlocal$: This case is
  analogous to Case 1; for completeness, we spell it out. We have
  $\Sigma_{\ttdlocal}(n_{\ttdlocal}) = \Tau_{\ttdlocal}(n_{\ttdlocal}) -
  1$, $\delta_{\ttdlocal} = \gamma_{\ttdlocal} - 1$, and $b_{\ttdlocal} =
  c_{\ttdlocal} - 1$. Using the IH, we get
  $\Sigma_{\ttdlocal}(n_{\ttdlocal}) = n_{\ttdlocal}
  \maxplus[c_{\ttdlocal}](\delta_{\ttdlocal} + 1) - 1$.
  \begin{itemize}

  \item If $n_{\ttdlocal} + \delta_{\ttdlocal} + 1 \atl c_{\ttdlocal}$,
    then $n_{\ttdlocal} \maxplus[c_{\ttdlocal}](\delta_{\ttdlocal} + 1) - 1
    = n_{\ttdlocal} + \delta_{\ttdlocal} = n_{\ttdlocal}
    \maxplus[b_{\ttdlocal}] \delta_{\ttdlocal}$ since $n_{\ttdlocal} +
    \delta_{\ttdlocal} \atl c_{\ttdlocal} - 1 = b_{\ttdlocal}$.

  \item If $n_{\ttdlocal} + \delta_{\ttdlocal} + 1 < c_{\ttdlocal}$, then
    $n_{\ttdlocal} \maxplus[c_{\ttdlocal}](\delta_{\ttdlocal} + 1) - 1 =
    c_{\ttdlocal} - 1 = b_{\ttdlocal} = n_{\ttdlocal}
    \maxplus[b_{\ttdlocal}] \delta_{\ttdlocal}$ since $n_{\ttdlocal} +
    \delta_{\ttdlocal} < c_{\ttdlocal} - 1 = b_{\ttdlocal}$.

  \end{itemize}

\item[Case 3:] $e_1 \in \ttdtranssexpand \setminus \ttdtranss$ and $l_1 =
  \ttdlocal$: Then $\Sigma_{\ttdlocal}(n_{\ttdlocal}) =
  \Tau_{\ttdlocal}(n_{\ttdlocal}) \maxminus 1 + 1$, $\delta_{\ttdlocal} =
  \gamma_{\ttdlocal}$, and $b_{\ttdlocal} = c_{\ttdlocal} \maxminus 1 + 1$.
  Using the IH, we get $\Sigma_{\ttdlocal}(n_{\ttdlocal}) = n_{\ttdlocal}
  \maxplus[c_{\ttdlocal}] \delta_{\ttdlocal} \maxminus 1 + 1$.
  \begin{itemize}

  \item If $c_{\ttdlocal} \atl 1$, then $b_{\ttdlocal} = c_{\ttdlocal}$, so
    $n_{\ttdlocal} \maxplus[c_{\ttdlocal}] \delta_{\ttdlocal} \atl
    c_{\ttdlocal} \atl 1$, hence $n_{\ttdlocal} \maxplus[c_{\ttdlocal}]
    \delta_{\ttdlocal} \maxminus 1 + 1 = n_{\ttdlocal}
    \maxplus[c_{\ttdlocal}] \delta_{\ttdlocal} = n_{\ttdlocal}
    \maxplus[b_{\ttdlocal}] \delta_{\ttdlocal}$.

  \item If $c_{\ttdlocal} = 0$, then $b_{\ttdlocal} = 1$.
    \begin{itemize}

    \item If $n_{\ttdlocal} + \delta_{\ttdlocal} \atl 1$, then
      $n_{\ttdlocal} \maxplus[c_{\ttdlocal}] \delta_{\ttdlocal} \maxminus 1
      + 1 = n_{\ttdlocal} + \delta_{\ttdlocal} \maxminus 1 + 1 =
      n_{\ttdlocal} + \delta_{\ttdlocal} = n_{\ttdlocal}
      \maxplus[b_{\ttdlocal}] \delta_{\ttdlocal}$.

    \item If $n_{\ttdlocal} + \delta_{\ttdlocal} \atm 0$, then
      $n_{\ttdlocal} \maxplus[c_{\ttdlocal}] \delta_{\ttdlocal} \maxminus 1
      + 1 = c_{\ttdlocal} \maxminus 1 + 1 = 1 = n_{\ttdlocal}
      \maxplus[b_{\ttdlocal}] \delta_{\ttdlocal}$.

    \end{itemize}

  \end{itemize}

\item[Case 4:] none of the above. In this case $e_1$ has no impact on the
  path summary generated by \algorithmref{Symbolically executing a path for
    local state l}. Thus, $\Sigma_{\ttdlocal}(n_{\ttdlocal}) =
  \Tau_{\ttdlocal}(n_{\ttdlocal})$; in particular we have $b_{\ttdlocal} =
  c_{\ttdlocal}$ and $\delta_{\ttdlocal} = \gamma_{\ttdlocal}$. Further,
  $\Sigma_{\ttdlocal}(n_{\ttdlocal}) = \Tau_{\ttdlocal}(n_{\ttdlocal})
  \symbolcomment{IH} = n_{\ttdlocal} \maxplus[c_{\ttdlocal}]
  \gamma_{\ttdlocal} = n_{\ttdlocal} \maxplus[b_{\ttdlocal}]
  \delta_{\ttdlocal}$.\eop

\end{description}

We now turn to the main goal of this section, the proof of
\theoremref{counter values}. We repeat it here for convenience,
\emphasize{except} that, applying \lemmaref{counter values single
  iteration}, we replace term $n_{\ttdlocal} \maxplus[b_{\ttdlocal}]
\delta_{\ttdlocal}$ in the original theorem formulation by
$\Sigma_{\ttdlocal}(n_{\ttdlocal})$, which simplifies the proof.
\setcounter{DEF}{\thetheoremCounterValues}
\begin{THE}
  Let superscript $\iterate{}$ denote $\iterator$ function applications.
  Then, for $\iterator \atl 1$,
  \begin{equation}
    \label{equation: proof: counter values}
    \iterate{\Sigma_{\ttdlocal}}(n_{\ttdlocal}) = \Sigma_{\ttdlocal}(n_{\ttdlocal}) \maxplus[b_{\ttdlocal}] (\iterator-1) \cdot \delta_{\ttdlocal} \ .
  \end{equation}
\end{THE}
\Proof: by induction on $\iterator$. For $\iterator=1$, the right-hand side
(rhs) of \equationref{proof: counter values} equals
$\Sigma_{\ttdlocal}(n_{\ttdlocal}) \maxplus[b_{\ttdlocal}] 0 =
\Sigma_{\ttdlocal}(n_{\ttdlocal})$ since $\Sigma_{\ttdlocal}(n_{\ttdlocal})
+ 0 = \Sigma_{\ttdlocal}(n_{\ttdlocal}) \atl b_{\ttdlocal}$ by
\lemmaref{counter values single iteration}.

Now suppose \equationref{proof: counter values} holds. For the inductive
step we obtain:
\begin{eqnarray}
  \iterate[(\iterator+1)]{\Sigma_{\ttdlocal}}(n_{\ttdlocal}) &                                                                   = & \Sigma_{\ttdlocal}(\iterate{\Sigma_{\ttdlocal}}(n_{\ttdlocal})) \nonumber \\
                                                    & \symbolcomment{IH}                                                = & \Sigma_{\ttdlocal}(\Sigma_{\ttdlocal}(n_{\ttdlocal}) \maxplus[b_{\ttdlocal}] (\iterator-1) \cdot \delta_{\ttdlocal}) \nonumber \\
                                                    & \symbolcomment{Lem.~\lemmaref[]{counter values single iteration}} = & (\Sigma_{\ttdlocal}(n_{\ttdlocal}) \maxplus[b_{\ttdlocal}] (\iterator-1) \cdot \delta_{\ttdlocal}) \maxplus[b_{\ttdlocal}] \delta_{\ttdlocal} \ . \label{equation: proof: counter values aux}
\end{eqnarray}
We now distinguish three cases ($\explain[]{\ldots}$ below contains proof
step justifications):

(1) If $\delta_{\ttdlocal} \atl 0$:
\[
\begin{array}{cl}
      & \mbox{\equationref{proof: counter values aux}} \\
  = \ & \explain{$(\iterator-1) \cdot \delta_{\ttdlocal} \atl 0$, $\Sigma_{\ttdlocal}(n_{\ttdlocal}) \atl b_{\ttdlocal}$, hence $\Sigma_{\ttdlocal}(n_{\ttdlocal}) + (\iterator-1) \cdot \delta_{\ttdlocal} \atl b_{\ttdlocal}$} \\
      & (\Sigma_{\ttdlocal}(n_{\ttdlocal}) + (\iterator-1) \cdot  \delta_{\ttdlocal}) \maxplus[b_{\ttdlocal}] \delta_{\ttdlocal} \\
  =   & \explain{$\delta_{\ttdlocal} \atl 0$} \\
      & (\Sigma_{\ttdlocal}(n_{\ttdlocal}) + (\iterator-1) \cdot \delta_{\ttdlocal}) + \delta_{\ttdlocal} \\
  = \\
      & \Sigma_{\ttdlocal}(n_{\ttdlocal}) + \iterator \cdot \delta_{\ttdlocal} \\
  =   & \explain{$\Sigma_{\ttdlocal}(n_{\ttdlocal}) + \iterator \cdot \delta_{\ttdlocal} \atl b_{\ttdlocal}$} \\
      & \Sigma_{\ttdlocal}(n_{\ttdlocal}) \maxplus[b_{\ttdlocal}] \iterator \cdot \delta_{\ttdlocal} \ ,
\end{array}
\]
the final expression being the rhs of \equationref{proof: counter values},
for $\iterator$ replaced by $\iterator+1$.

(2) If $\delta_{\ttdlocal} < 0$ and $\Sigma_{\ttdlocal}(n_{\ttdlocal}) + (\iterator-1) \cdot \delta_{\ttdlocal} <
b_{\ttdlocal}$, then also $\Sigma_{\ttdlocal}(n_{\ttdlocal}) + \iterator \cdot \delta_{\ttdlocal} < b_{\ttdlocal}$, and:
\[
\begin{array}{cl}
      & \mbox{\equationref{proof: counter values aux}} \\
  = \ & \explain{$\Sigma_{\ttdlocal}(n_{\ttdlocal}) + (\iterator-1) \cdot \delta_{\ttdlocal} < b_{\ttdlocal}$} \\
      & b_{\ttdlocal} \maxplus[b_{\ttdlocal}] \delta_{\ttdlocal} \\
  =   & \explain{$\delta_{\ttdlocal} < 0$} \\
      & b_{\ttdlocal} \\
  =   & \explain{$\Sigma_{\ttdlocal}(n_{\ttdlocal}) + \iterator \cdot \delta_{\ttdlocal} < b_{\ttdlocal}$} \\
      & \Sigma_{\ttdlocal}(n_{\ttdlocal}) \maxplus[b_{\ttdlocal}] \iterator \cdot \delta_{\ttdlocal} \ .
\end{array}
\]

(3) If finally $\delta_{\ttdlocal} < 0$ and
$\Sigma_{\ttdlocal}(n_{\ttdlocal}) + (\iterator-1) \cdot \delta_{\ttdlocal}
\atl b_{\ttdlocal}$, then \equationref{proof: counter values aux} reduces
to $(\Sigma_{\ttdlocal}(n_{\ttdlocal}) + (\iterator-1) \cdot
\delta_{\ttdlocal}) \maxplus[b_{\ttdlocal}] \delta_{\ttdlocal}$. To get an
overview of what we need to prove, let
\[
\begin{array}{rclcrcl}
  X & = & \Sigma_{\ttdlocal}(n_{\ttdlocal}) + (\iterator-1) \cdot \delta_{\ttdlocal} \ , & \quad & X' & = & \Sigma_{\ttdlocal}(n_{\ttdlocal}) \ , \\
  Y & = & \delta_{\ttdlocal}                                   \ , &       & Y' & = & \iterator \cdot \delta_{\ttdlocal} \ .
\end{array}
\]
Then (the reduced) \equationref{proof: counter values aux} equals $X
\maxplus[b_{\ttdlocal}] Y$, and the rhs of \equationref{proof: counter
  values} equals $X' \maxplus[b_{\ttdlocal}] Y'$. Further, observe that $X
+ Y = X' + Y'$. This implies that $X \maxplus[b_{\ttdlocal}] Y = X'
\maxplus[b_{\ttdlocal}] Y'$, which follows immediately by distinguishing
whether $X+Y \atl b_{\ttdlocal}$ or not. The equality $X
\maxplus[b_{\ttdlocal}] Y = X' \maxplus[b_{\ttdlocal}] Y'$ is what we
needed to prove.\eop

} 

\end{document}